\documentclass[usenatbib]{basi}
\usepackage[T1]{fontenc}
\usepackage[british]{babel}
\usepackage[varg]{txfonts}
\usepackage{graphicx} 
\usepackage{epstopdf}
\usepackage{url}
\begin{document}
\title[Red population of Abell 1314]{Red population of Abell 1314 : A rest-frame narrowband photometric evolutionary analysis}
%            \texttt{http://www.mrao.cam.ac.uk/\string~dag/}
%
\author[Yuvraj Harsha Sreedhar]%
       {Yuvraj Harsha Sreedhar $^{1}$\thanks{E-mail:
yuvrajharsha@gmail.com (YHS)} \\
$^1$ University of Vienna, Institute of Astronomy, T\"urkenschanzstra{\ss}e 17, A-1180 Vienna, Austria}

\pubyear{2014}
\volume{42}
\pagerange{\pageref{firstpage}--\pageref{lastpage}}
%\status{submitted}

\date{Received 2014 April 24; accepted 2014 September 30}

\maketitle
%------------------------------------------------------------------------------%
% abstract and keywords                                                        %
%------------------------------------------------------------------------------%
\label{firstpage}

\begin{abstract}
Red sequence galaxies form with an intense burst of star formation in the early universe to evolve passively into massive, metal rich, old galaxies at \textit{z} $\sim$ 0. But Abell 1314 (\textit{z}=0.034) is found to host almost all red sequence galaxy members -- identified using the \textit{mz} index, classified using the Principle Component Analysis technique and SDSS colour correlations -- some of which show properties of low-mass, star forming, and metal rich galaxies. The variably spread Intra-Cluster Medium (ICM) near the core forms a vital part in influencing the evolution of these members. To study their evolution, I correlated different parameters of the rest-frame narrowband photometry and the derived luminosity-weighted mean Single Stellar Population model ages and metallicities. 

The study finds the member galaxies evolve differently in three different sections of the cluster: 1. the region of $\leq$ 200 kpc hosts passively evolving old, massive systems which accumulate mass by dry, minor mergers, 2. the zone between 200-500 kpc shows stripped systems (or in the process of being gas stripped) by ram pressure with moderate star formation history, 3. the outer regions ($\geq$ 500 kpc) show low-mass red objects with blue, star forming Butcher-Oemler galaxy like colours. This sort of environmental condition is known to harbour hybrid systems, like, the pseudo bulges, blue sequence E/S0 and Butcher-Oemler like satellite cluster galaxies. Overall, the cluster is found to be poor, quiescent with galaxies to have formed by the monolithic structure formation in the early universe and are now evolving with mergers and gas stripping processes by ram pressure.

\end{abstract}


%------------------------------------------------------------------------------%
% main text of the paper, using \section, \subsection, \subsubsection          %
%------------------------------------------------------------------------------%
\section{Introduction}\label{s:intro}

The red-blue classification scheme (Strateva et al. 2001; Bell et al. 2004) shows a bimodal division of galaxies with varying properties in their size, surface density, concentration, formation and evolution over the cosmic time. The red sequence galaxies (earlier than Sa types) consist of passive stellar populations at \textit{z} $\leq$ 1, with low to no star formation, are metal rich, massive and old. Whereas, the blue sequence galaxies (Sd-Sm) are primarily star-forming, with a wide range of metallicity values and are comparatively low-mass and younger than the red sequence counterparts. These properties of the two groups relate to various phenomena, like downsizing, Butcher-Oemler effect (Butcher \& Oemler 1978), changes in age, metallicity, star formation rate, clustering, kinematics, to name a few. 

This scheme also separates galaxies based on their bulge to disk (B/D) ratio. Such that, classical bulges, structurally similar to ellipticals with hosting old stellar populations, are observed to have formed through fast monolithic collapse (Eggen et al. 1962) and by processes of merging, harassment, stripping of disks (Lorenzo et al. 2014) while rotating in clusters (Kormendy \& Kennicutt 2004). Pseudo bulges appear relatively flat (as edge on) to have formed through slow accretion of gas and dust, corresponding to `secular' process (Kormendy \& Kennicutt 2004) in field showing the presence of disks, rings and bars. The contained gas is known to channel (Schwarz et al. 1981) to and fro from the bar and central bulge leading to star formation (fairly recent) and hence, hosts young stars. Both these types of bulge systems represent their formation through different evolutionary times (Drory \& Fischer 2007). Although several authors (Kormendy 1979a,b; 1981, 1982a; Norman 1984; Pfenniger \& Norman 1990; Combes 1991; Buta 1995, 2000; Kormendy \& Gebhardt 2001; Athanassoula 2002; Gadotti 2009a) have been studying the properties of these hybrid bulges, but their formation and evolution still present many open questions which needs to be analysed in-depth.

But in past years, this classification scheme has also shown evidences of overlapping of red and blue population properties, such as the blue sequence E/S0s (Kannappan et al. 2009; Wei et al. 2010). The blue sequence E/S0s are the galaxies which are found in the late-type galaxy section of the colour-stellar mass diagram. These hybrid systems represent the early red objects, but with blue object properties; based on their stellar mass, these blue E/S0s  formed by mergers are now (at \textit{z}$\sim$0) showing properties of star formation, disk building while evolving in moderate density environments. Learning about the processes which form these blue-sequence E/S0s to probably evolve into large spiral disks and escape from the dwarf irregular regime, would make a fascinating study (Kannappan et al. 2009).

In conjunction to the studies of by Drory \& Fischer (2007) and Kannappan et al. (2009), alongside the main objective of understanding the formation and evolution of these red sequence galaxies at different redshifts and environments, I report photometric results of galaxies in cluster Abell 1314 using the rest-frame narrowband colours. Since for a galactic evolutionary analysis, parameter sensitive indices play a vital role for identifying stellar populations based on their temperature, metallicity, age, star formation while having optimal spatial resolution, the technique of modified Str\"omgren (Rakos et al. 1988, 1990, 1991, 1995) photometry is found to be suitable for this study. The broadband photometry although offers a good spatial resolution but is affected by the poor spectral resolution that is required for precise measurement of different evolutionary parameters, hence are greatly influenced by the age-metallicity degeneracy (Worthey 1994). While, on the other hand, spectroscopy, in spite of providing high spectral resolution, the technique fails to break the degeneracy as different elemental abundances confuses the evolutionary picture (Rakos et al. 2001a, Sreedhar et al 2012).

This cluster, at \textit{z} =0.034, forms an ideal target to serve both these studies (evolution of galaxies in A1314 and exploration of blue-sequence E/S0s) with a richness class III and Bautz-Morgan class 0; from the reddening impervious \textit{mz} index (Fiala et al. 1986) of the modified Str\"omgren photometry and the Principle Component Analysis (PCA; Steindling et al. 2001), I find, the cluster hosts almost all red-sequence galaxies as its members up to a distance of $\sim$1 Mpc. To base the current study primarily upon modified Str\"omgren photometry, I use, as much as possible, rest frame colour indices and their properties for galaxy classification, removal of fore- and background objects, and age-metallicity determination.

The non-uniformly spread Inter-Cluster Medium (ICM) shows its presence by the soft excess, low X-ray emission (at 2-6 keV band with $L_x$ $\le$ 7 $\times$ $10^{44}$ erg $s^{-1}$) and is known to influence the evolution of member galaxies by ram pressure stripping. The effect of ram pressure on the star formation rate of a radially infalling disk galaxy in a cluster has been theoretically studied 
Fujita \& Nagashima (1998) and Kenji (2013). The ram pressure stripping can enhance, reduce and quench star formation depending upon on the group/cluster halo masses, pericentre distances of orbits, inclination angles of disks with respect to the direction of their orbits. 

Though, the cluster was examined earlier in radio by Vall\'ee \& Wilson (1976), Wilson \& Vall\'ee (1977), Vall\'ee \& Roger (1987), exposed in X-ray using Einstein Imaging Proportional Counter (Jones \& Forman 1999), and in optical (Flin et al. 1995; Lauer et al. 2007) for a few central members (IC 708, IC 711, IC 709) .  In the current study, the influence on galaxy evolution in A1314 for the first time is found by correlating various properties (colour evolution, mass, ages, metallicities, star formation history, number density across the radial distance) determined by our rest-frame narrowband photometric method in the UV-Optical waveband. 

The present study is arranged as follows: Section 2: Radio and X-ray surveys; 
Section 3: Observations; Section 4: Results; Section 5: Discussions which cover topics of cluster membership criteria, spectrophotometric classification, morphology inspection using SDSS colour indices (concentration index and \textit{u-r}), multi-colour diagrams, colour-magnitude diagram, Butcher-Oemler effect, radial galaxy density, age, metallicity determination, PCA supported SSP models,  luminosity weighted mean age and metallicity analysis, and their correlation with cluster radial density; and Section 6: Conclusions. 

%------------------------------------------------------------------------------%
% example figures to illustrate various styles...
%------------------------------------------------------------------------------%

\begin{figure}
\includegraphics[scale=0.6]{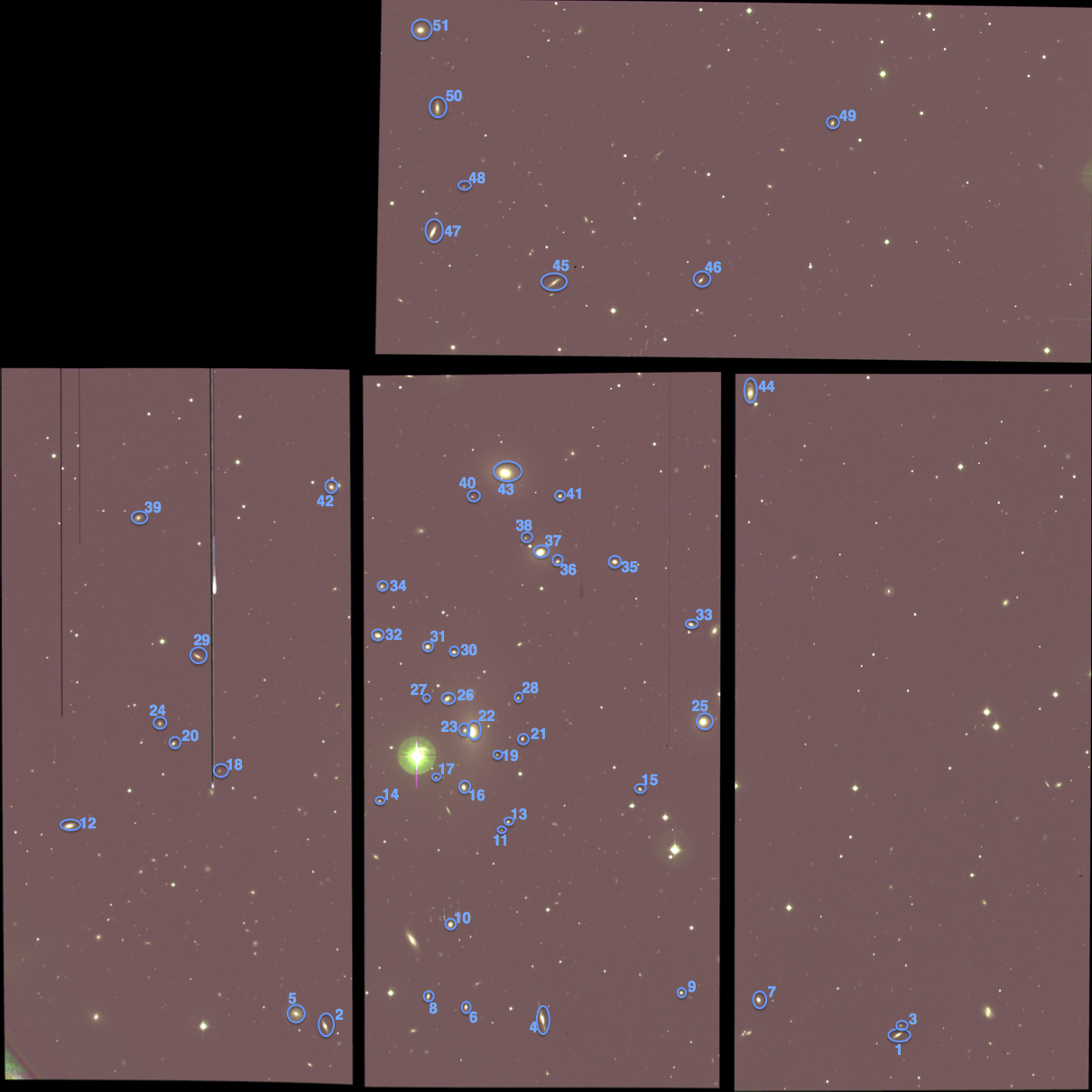}
%\vspace{100pt}
% \centering  % this centres figure in column
 %\includegraphics{cc.ps}
  \caption{Multi-band colour image of A1314 (N is pointing left and E is to the bottom of the page): Combined colour image of \textit{vz, bz, yz} filters with member galaxies, found using the \textit{mz} index, are marked in blue circles and numbered with galaxy serial numbers that are tabulated in Table 2. One should note, that the cluster membership is not complete with the \textit{mz} index, but adequate in the evolutionary sense of A1314 galaxies within the photometric limits of magnitude, resolution and cluster centric distance to avoid any contamination and fallacy in the measurements.}
\end{figure}

%------------------------------------------------------------------------------%
\section[]{Radio and X-ray insights of A1314}

The X-ray and radio observations of galaxy clusters can greatly contribute to our understanding of their formation and evolution; they reveal the presence of X-ray emitting hot gas (via thermal bremsstrahlung emission), ICM, interacting diffuse radio sources (like, radio halos and relics) with ICM to emit relativistic electrons and produce large scale magnetic fields. Their presence reflect the massive accumulation of dark matter halo (Popesso et al. 2004). When these X-ray and radio measurements are combined with spectrophotometric measurements in UV-optical wavelengths, a lot can be learnt about the baryonic components and their properties relating to the structure formation in the Universe.

Although Abell 1314 has been observed in both wavelengths, but it has gained more popularity in the radio wavelength because of the two radio bright galaxies -- IC 708 and IC 711 -- which on interaction with the ICM have become the head-tail or the narrow-angle tailed (NAT) radio sources (O'Dea \& Owen 1985). IC 711 has one of the longest tails ever detected in the known Universe, extending up to 630 kpc (Vall\'ee \& Roger 1987; Bliton et al. 1998). The unusual bend in this tail is formed due to the ram pressure of the high-velocity host galaxy moving through the dense ICM forms sub-peaks near the end and a double structure just a few arc seconds before the nucleus of the galaxy (Vall\'ee \& Wilson 1976). 

The most astonishing aspect of the radio pair comes from the relativistic electrons in IC 711 tail, traveling such great lengths ($\sim$630 kpc), which cannot originate in the nucleus of IC 711 because of the energy loss. Wilson \& Vall\'ee (1977) believe that the tail is being driven through the surrounding hot gas from the ICM by the influence of gravitational forces. They suggest either the tail is buoyant with bubbles of plasma rising in the gravitation field, dominated by IC 708, or the tail is heavy and falling towards the cluster centre. By these predicted models, the terminal velocities of the northern part of the tail could reach speeds of a few hundred km $s^{-1}$, and that could account for the bending. 

Bonamente et al. (2002) find, within large statistical error bars, the absence of cooler gas at the cluster centre to convey that the cluster may host soft excess emission throughout the central 500 kpc, where the cluster emission is detected. Wilson \& Vall\'ee (1977)  find the X-ray luminosity in the 2-6 keV band $L_x$ $\le$ 7 $\times$ $10^{44}$ erg $s^{-1}$. Flin et al. (1995) view the cluster as very clumpy with elongated X-ray emission and a strong X-ray centroid shift. This elongation is also present in the galaxy distribution that shows a definite ellipticity, and is oriented in the E-W direction.

\section{Observations}
The photometry for this project was obtained on 2 May 2005 on the 2.54 m INT, La Palma with the prime-focus imager, that has a field of view (FOV) of $20'$ $\times$ $40'$ corresponded to almost 1 Mpc of A1314; the wide field camera consists of four 2k $\times$ 4k mosaic CCDs and a plate scale of $0^{\prime\prime}.33$  pixel$^{-1}$. The cluster was observed through 4 sets of 600 s exposures, making a total of 2400 s per filter. Since a part of the night was affected by the dusty winds from the Sahara, the calibration was obtained using the average of five nights of spectrophotometric standards, instead of just one. Fig. 1 shows the coloured, combined image of \textit{vz, bz, yz} filters, with the member galaxies marked in blue circles found using \textit{mz} index (discussed in \S 4.1 and \S 4.2). 

The modified Str\"omgren filter system used for the observation of the Abell 1314 is defined in Rakos \& Schombert (1995) and Sreedhar et al. (2012) and the references therein. In brief, the modified Str\"omgren filter system (\textit{uz, vz, bz, yz}) is the redshifted Str\"omgren (\textit{u, v, b, y}) filters. The \textit{uz, vz, bz, yz} system covers three regions in the near-UV and blue portion of the spectrum, that makes it a powerful tool for the investigation of stellar populations in SSPs, such as star clusters, or composite systems (galaxies). The first region is long ward of 4600 \AA\, where the influence of absorption lines is small. This is characteristic of the \textit{bz} and \textit{yz} filters at $\lambda_{eff}$ = 4675 \AA\ and 5500 \AA, respectively, which produce a temperature colour index (\textit{bz-yz}). The second region is a band short ward of 4600 \AA, but above the Balmer discontinuity. This region is strongly influenced by metal absorption lines (i.e., Fe and CN), particularly for spectral classes F-M, which dominate the contribution of light in old stellar populations. This region is exploited by the \textit{vz} filter ($\lambda_{eff}$=4100 \AA). The third region is a band short ward of the Balmer discontinuity or below the effective limit of crowding of the Balmer absorption lines. This region is explored by the \textit{uz} filter ($\lambda_{eff}$ = 3500 \AA). All the filters are sufficiently narrow (FWHM = 200 \AA) to sample regions of the spectrum unique to the various physical processes of star formation and metallicity (see Rakos et al. 2001a for a fuller description of the colour system and its behaviour for different populations). For the filter transmission curves and the bracketed spectral regions, the readers are referred to Fig. 1 of Rakos et al. (1995) and Sreedhar et al. (2012). 

The reduction procedure has been published in Rakos et al. (1996). The photometric system is based on the theoretical transmission curves of filters (which can be obtained upon request from the author) and the spectra of the spectrophotometric standards are published by Massey \& Gronwall (1990) and Hamuy et al. (1994). The convolution of the transmission curves and the spectra of the standard stars, which produce theoretical flux values for colour indices, are corrected for all light losses in the equipment and the specific sensitivity of the CCD camera. Magnitudes are measured on the resultant co-added images using standard IRAF procedures and for brighter objects are based on metric apertures set at 32 kpc for cosmological parameters of $H_o$ = 75 km $s^{-1}$ $Mpc^{-1}$ and the benchmark cosmology ($\Omega_m$ = 0.3, $\Omega_\Lambda$ = 0.7). Stellar mass is calculated from $M_{5500}$ luminosity  assuming M/L of 3 (Bender et al. 1992). For the fainter objects, the apertures are adapted to deliver the best possible signal-to-noise ratio, but always maintaining the same aperture for all four filters.

For the galactic extinction \textit{E(B-V)} = 0.0181 mag for A1314 found from the NASA/IPAC Infrared Science Archive (IRSA), corresponding modified St\"omgren colours and typical photometric errors, for observational limiting magnitude of 14.3 $\leq$ $m_{5500}$ $\leq$ 19.2 (-16.4 $<$ $M_{5500}$ $<$ -21.3), are shown in Table 1. The effect of these low photometric uncertainties is negligible and insignificant on the derived ages and metallicities (within the specified limits).

\begin{table}
\caption{Modified Str\"omgren Photometric Precision}
\small
\centering
%\begin{}
\begin{tabular}{lll}
\hline  \\
Colour  & Galactic & Photometric  \\ 
 indices &  extinction & errors \\ 
  & mag & mag \\ 
 \hline \\
\textit{uz-vz} & 0.01 & 0.07 \\
\textit{bz-yz} & 0.01 & 0.02 \\
\textit{vz-yz} & 0.03 & 0.05 \\
& \multicolumn{1}{l}{} & \multicolumn{1}{l}{}  \\ 
\hline \\
\end{tabular}
%\end{}
\end{table}
%\newpage

\section{Discussions}

\subsection{Cluster Membership Criteria}

The \textit{mz} (=\textit{(vz-bz)-(bz-yz)}) index offers a combined merit--of being impervious to reddening, redshifted to the target cluster and shows almost linear change in colour with redshift (between the -0.5 $\leq$ \textit{z} $\leq$ 0.09 range; Fiala et al. 1986)--these properties can be exploited to discriminate fore- and background objects (including stars) by considering the amount of velocity deviation of an object by the cluster. Large velocity deviations from the mean cluster redshift can introduce discordant values, but as long as the velocity deviation remains within 2000 km $s^{-1}$ (which is in most general case at low \textit{z}), the individual velocities of member galaxies will contribute only a very small effect on the mean colour index (Rakos et al. 1991).

Whilst, this property is well documented by Fiala et al. (1986), Rakos et al. (1988) for standard ellipticals for clusters up to \textit{z}=0.2, (where Bruzual (1983) finds \textit{mz} to be an ideal index to discriminate members from field up to \textit{z}=0.5), Rakos et al. (1996) have successfully used the same index to understand galaxies in A2317 at \textit{z}=0.211. Also, the practical application of the index's sensitivity towards identifying and studying the late-type, blue sequence galaxies are shown in Rakos et al. (1995, 1996), Butcher-Oemler property in A2317 (Rakos et al. 1997). These works convey the ability of this index towards identifying cluster members of both red and blue sequence galaxies. However, to examine, again, the two-colour efficiency of \textit{mz}, the identified members are compared with a more \textit{robust and efficient} four-colour PCA technique (Steindling et al. 2001). But its important to note that all the colours used in either techniques are based on the same modified Str\"omgren filter system.  

Besides being \textit{efficient} in identifying cluster members to much fainter ($\sim$ 20 mag) limits, this four-colour PCA technique is \textit{robust} enough to classify galaxies (discussed in the next section) into different galaxy types. This PCA technique, developed using the synthetic Str\"omgren colours for a set of 150 spectra of nearby galaxies covering all possible morphological types, is able to separate galaxies in to a limited region of the three-dimensional colour space.  Since real galaxy spectra of various types including dust-enshrouded cluster member were used as templates, and not models, there is no worry about different constituents of model galaxies and their evolutionary bias (Steindling et al. 2001). The efficiency of this rejection mechanism is characterised by the selected maximum deviation sigma between any measured colour from the same colour of the best template galaxy. Note that our filters are narrow enough to obtain data of sufficient spectral resolution for identification and classification purposes, but sufficiently wide, such that, the cluster's velocity dispersion does not affect the colours.

The \textit{mz} index finds 51 member galaxies of this low-\textit{z} A1314 cluster in the 20' $\times$ 40' FOV. But by using PCA technique, we find 89, of which 42 are same objects detected by \textit{mz} index. While in the rest of the cases, 21 are found to be too faint ($>$ 19.2 mag) to get proper scientific information, 20 objects lie close to another that they are difficult to spatially resolve for correct aperture photometry, 1 is photometrically saturated for proper colour measurement and 5 were found to be possibly valid member galaxies. Although this would convey that the PCA technique is much superior and \textit{efficient} in identifying cluster members than \textit{mz} index, but since most of the galaxies detected by the \textit{mz} index are found to be common to those using the PCA technique, while remaining others are hampered for proper scientific investigation (because of faintness, close spatial proximity to one another, saturation etc.), therefore I examine the evolutionary analysis of those detected \textit{mz}. Because of the aforementioned issues, the sample of the cluster members is not complete, and recommend an investigation of the hampered galaxies (detected by the PCA technique) using high spatially resolved imaging.

\subsection{Spectrophotometric Classification using PCA Technique}

Earlier, Rakos et al. (1996) demonstrated a photometric galaxy classification scheme using two-colours and \textit{mz} diagrams by carefully studying 150 high resolution and high S/N spectra to generate narrowband colours of different types of galaxies. The system allows  classification of galaxies into four simple types (Ellipticals/S0s, Spirals/Irregulars, Seyferts, and Starbursts) in a purely spectrophotometric manner.

This method is enhanced by Steindling et al. (2001) using the \textit{robust} PCA technique by studying a collection of large aperture spectra. This PCA technique is applied to the rest-frame observations of Coma (Odell et al. 2002) to define and present the classes of galaxies based on their narrowband colours. The division of galaxies is made such that the first principal component axis (PC1) divides into four subdivisions based on mean past star formation rate at different extinction values: E (passive, red objects), S (star formation rates equivalent to a disk galaxy), S- (transition between E and S; also referred as transition galaxies), and S+ (starburst objects). 

\begin{figure}
%\vspace{100pt}
 \centering  % this centres figure in column
  \includegraphics[scale=0.55]{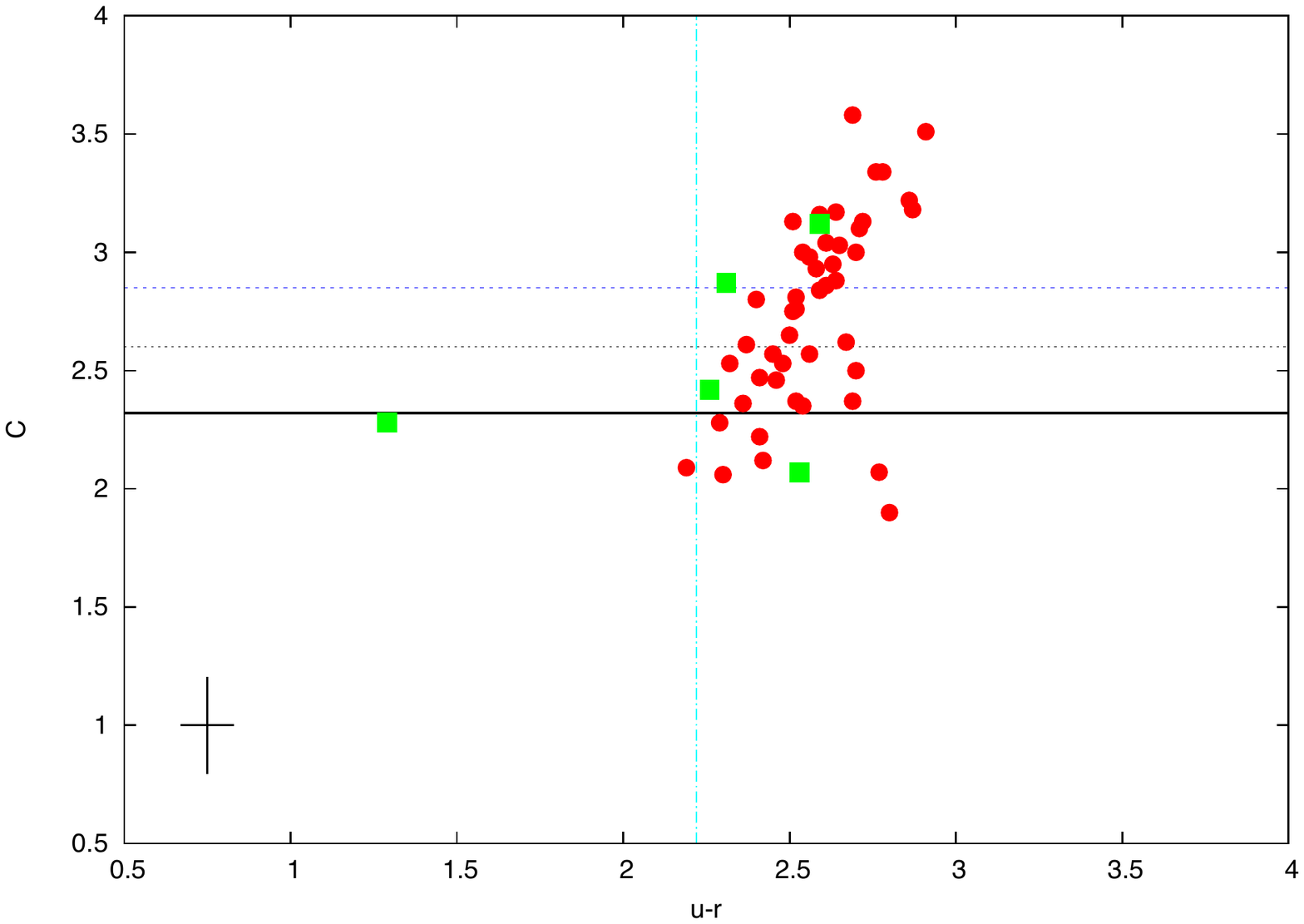}
  \caption{Member galaxy correlation with Concentration Index (C) and SDSS \textit{u-r} colour index, showing a linear relation, for morphological classification : The member galaxies of A1314 found using \textit{mz} are shown in red circles and those using with the four-colour technique are displayed in green squares. Galaxies below the 2.32 C mark and left of \textit{u-r} 2.2 mark are classified as late-types, but here almost all points are found to lie in the early-type section of the plot. The purple dashed line (C=2.85) is adopted by Nakamura et al.  (2003), the black dotted line at 2.6 by Shimasaku et al. (2001) and solid line at 2.32 is by Blanton et al. (2001) and the vertical line at the SDSS \textit{u-r} line at 2.22 as the limits of separating the early- from the late-types. Error bars indicate the standard deviation of the mean of C and \textit{u-r} values of member galaxies.}
\end{figure}

By comparing spectral energy distribution (SED) of models and observations of nearby galaxies (Rakos et al. 1996), these divisions along the PC1 axis are drawn, such that, S galaxies correspond to those systems with Spiral star formation rates (approximately 1 $M_{\odot} yr^{-1}$) and S+ galaxies correspond to Starburst rates (approximately 10 $M_{\odot} yr^{-1}$). Since these divisions are determined by continuum colours versus spectral lines, they do not represent the current star formation rate, but rather the mean star formation rate averaged over the last few gigayears, as reflected by the optical emission by the dominant stellar population. 

The red E systems display colours with no evidence of star formation in the last 5 Gyr. The transition objects, S-, represent the fact that there is no sharp division between the E and S class. These objects display slightly bluer colours (statistically) from the passive E class; however, the difference could be due to a recent, low-level burst of star formation or a later epoch of galaxy formation or an extended phase of early star formation or even lower mean metallicity (i.e., the colour-magnitude effect).

In addition to the classification by colour, the technique allows to separate objects with signatures of non-thermal continuum (active galactic nuclei) under the categories of A+, A, and A- based on their PC2 values. It is important to remember that these classifications are based solely on the principal components (Rakos et al. 2007). 

Surprisingly, almost all member galaxies (which are identified using \textit{mz} index) are classified as red sequence using the PCA technique; among these member galaxies, there 38 are ellipticals (E, passive, red objects) to constitute 75\%, and 13 are transition galaxies or S- types (S0s with transitional properties between ellipticals and spirals) forming the other 25\% of cluster members. Since this classification scheme is based on colour (of the dominant stellar population in a galaxy) than the morphology (although being akin to their morphology), there could be some misclassification possible on the basis of the galactic shape, especially for those galaxies which appear diffused and fuzzy in our narrowband images; hence I find 'almost' all galaxies as red sequence, if not all.

\begin{figure}
\includegraphics[scale=0.5]{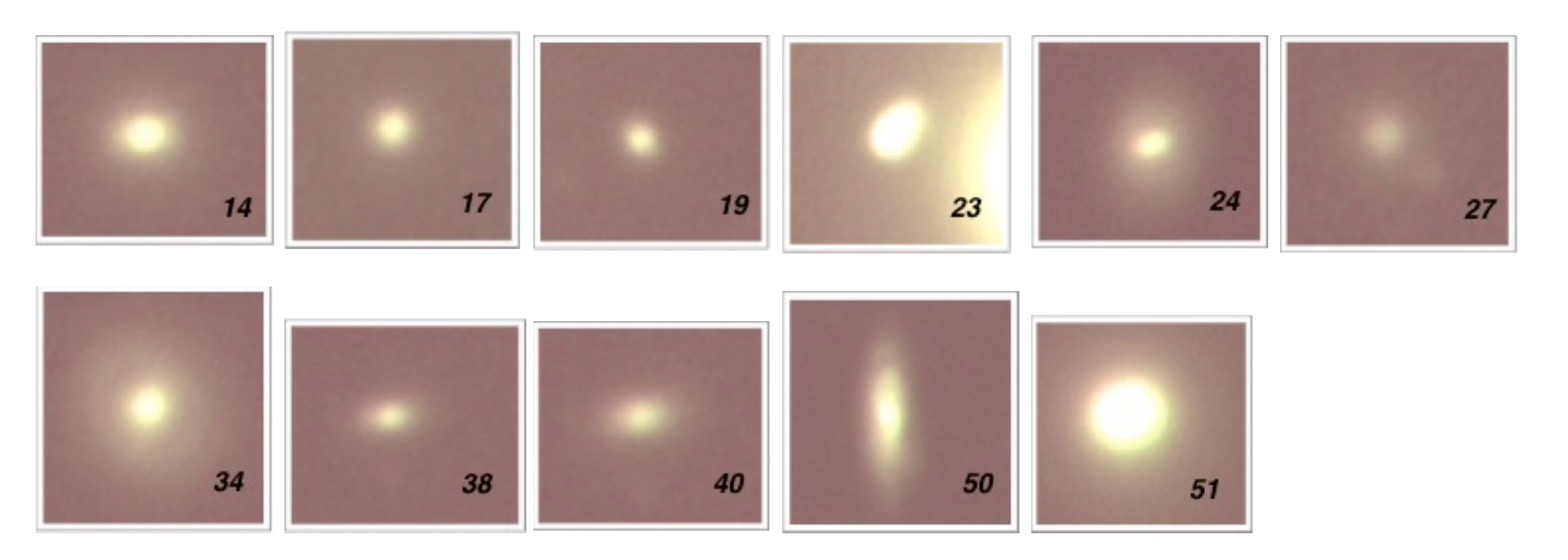}
%\vspace{100pt}
% \centering  % this centres figure in column
 %\includegraphics{cc.ps}
  \caption{The figure displays A1314 member galaxies (found using \textit{mz} index) with Concentration Index C $<$ 2.3 (object no. 19, 23, 24, 34, 38, 40, 50;  as per individual serial numbers listed in Table.1), \textit{g} $\geq$ 19.4 mag (14, 17, 27) and one with no SDSS data (51). Although galaxies with C $<$ 2.3 are classified as the late-types, but these diffused galaxies resemble the early-types.}
\end{figure}

\subsection{Morphological Inspection using SDSS Photometric Indices}

Since A1314 is a low \textit{z} cluster, data for almost all member galaxies (except one) are  available at the Sloan Digital Sky Survey DR8 (SDSS hereon; York et al. 2001; Aihara et al. 2011) site. Of the many tools offered by the SDSS\footnote{\url{https://www.sdss3.org/dr8/software/}} on imaging and spectroscopy for a thorough inspection of galaxies, the concentration index (Shimasaku et al. 2001, or the inverse concentration index; Strateva et al. 2001) emerges as a suitable proxy technique for a morphological analysis of galaxies. It is the ratio of the radii containing 90\% and 50\% of the Petrosian (Petrosian 1976) \textit{r} galaxy light, \textit{C = $r_{90}$/$r_{50}$}, that is known to correlate with individual galaxy type. Since Bulge to disk ratios are found to show a large scatter in separating 
the late- and early-types, the concentration index forms much better suited tool, since it directly relates to the B/D (Kent 1985; Hashimoto \& Oemler Jr. 1998; Shimasaku et al. 2001; Graham \& Worley 2008; Deng 2013).

Bulge dominated galaxies like the ellipticals are expected to have larger concentration index than disk structures (Strateva et al. 2001). Readers are cautioned to not to confuse the SDSS magnitudes and colour index used in this subsection, alone, from the modified Str\"omgrem photometric indices employed in other sections of this article. The objective of this study is to find out the morphologies of A1314 member galaxies from a broadband photometric reference frame in contrast to the narrowband photometry.

Different authors have studied this parameter to find different agreeable values to suit their data and analysis. Shimasaku et al. (2001) found the visual morphological correlations to be strongest with the concentration index amongst other \textit{PHOTO}\footnote{A SDSS pipeline software. For more details, refer \url{http://www.sdss.org/dr7/algorithms/photometry.html}} parameters, like the surface brightness, colour, asymmetry. They adopt the C index of 2.6 from the optimisation of subsamples using spectroscopy and morphology, however, they recommend C=3.03 on the basis of bright samples with low contamination. Blanton et al. (2001) found the separation at an inverse concentration index of 0.43. Nakamura et al. (2003) find no noticeable trend of fainter early-type galaxies having softer cores; most of the data points fall below inverse concentration index, C $<$ 0.34, a typical value that divides early and late types, down to -19 mag. Their inverse C index of 0.35 separated early- and late-types, where the both samples had a completeness of 82\%. Park \& Choi (2005) convey the concentration index as a good and simple parameter for classifying galaxy morphologies, and their chosen value for inverse concentration index is 0.35 (C=2.86).

While there is still a widespread debate on the completeness and the reliability achieved by these different concentration indices for a clear distinction, Strateva et al. (2001) demonstrated a SDSS colour selection scheme for spectroscopic and morphological selected sample of early and late-types. This (\textit{u-r}) colour of the SDSS filters \textit{u, g, r, i, z} selection shows a much better reliability in separating the bulges and disk structures, such that, most early-types have \textit{u-r} $\geq$ 2.22 and the late-types are found to be lower than this value. Kelly \& Mckay (2004) recommend that the value of 2.2 is just a rough boundary for separating the two types of galaxies, while \textit{u-r} $\geq$ 2.4 offers a clearer borderline. 

I correlated the measured C values with the SDSS colour \textit{u-r} to illustrate in Fig. 2 for all A1314 galaxies but three: two of which exceeded the limiting magnitude \textit{g} $\geq$ 19.4 mag (J113458.11+490545.5, J113442.97+490608.3) and, one, that showed bizarrely high \textit{u-r} value of 5.5 (J113502.84+490731.7). The three horizontal lines show the adopted C values by Nakamura et al. (2003) in a purple dashed line at 2.85, black dotted line by Shimasaku et al. (2001) at 2.6 and at 2.32 with a solid black line by Blanton et al. (2001). The vertical line at 2.2 gives the SDSS \textit{u-r} colour boundary by Strateva et al. (2001). Also, notice, that (\textit{u-r}) shows a linear relation with the C.

The points in red are the \textit{mz} index chosen members, while 5 galaxies which are found using the four-colour PCA technique, just for comparison, are displayed in green squares. Almost all the points lie linearly (with a slope of 1.4) towards the early-type section of the plot, except a few points which fall below the border line proposed by Blanton et al. (2001). These galaxies, which lie below the C = 2.3 border line, on visual inspection, resemble the early-types while some look diffused to make a proper judgement on their morphological structures. A spatially deep high-resolution imaging can elaborate on their physical properties of these galaxies. Galaxies having C $\leq$ 2.3 along with those which have \textit{g} $\geq$ 19.4 mag and, the one, (J113234.81+490634.7) with no SDSS data are displayed in Fig. 3 with their respective serial numbers from Table 2 (showing the list of A1314 cluster members with their determined properties). In conclusion, the correlation of C with \textit{u-r} index supports the notion of almost all A1314 member galaxies are early-types, as inferred by the PCA classification scheme.

\begin{figure}
\includegraphics[scale=0.65, angle=0]{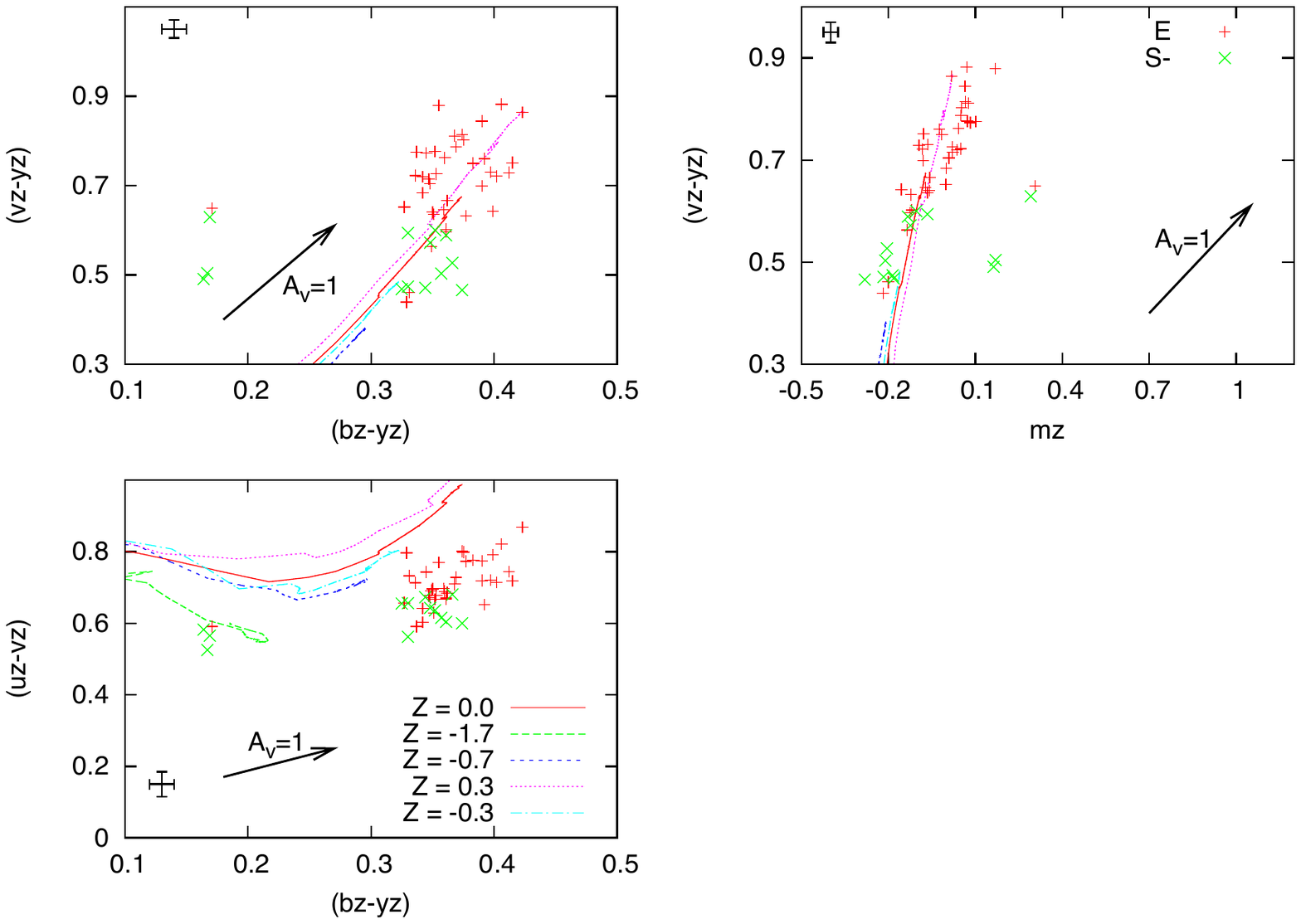}
\caption{Multicolour Diagram for A1314 Galaxies: All colour and \textit{mz [=(vz-bz)-(bz-yz)]} indices are correlated for cluster members, found by our spectrophotometric membership and galaxy classification scheme. The ellipticals are displayed by red plus signs, whereas the green crosses depict S-, transition galaxies. The SSP model tracks, for different ages and metallicities, from GALEV (Kotulla et al. 2009) passing through the cloud of points present a fair agreement with the modified Str\"omgren colours, considering the galaxies under investigation are mostly red, none with ongoing strong and complex star formation episodes, which are accounted by the superposition of different age/Z SSP model tracks. The error bars and the reddening vectors are also displayed for individual colour indices.}
\end{figure}

\subsection{Multicolour Diagrams}

Four narrowband colour indices form the basis for the current investigation of the cluster members; these correlations are displayed in Fig. 4, with the markings of red plus symbols for ellipticals and green crosses for the transition type (S-) galaxies. The respective colour error bars are displayed at the corners of individual plots and interstellar dust reddening vectors are shown for $A_{v}$ = 1. In addition, for comparison of model estimates, SSP  metallicity tracks from GALEV models (Kotulla et al. 2009) for various ages are also displayed and labelled. 

The (\textit{bz-yz}), (\textit{vz-yz}) multicolour plot displays interesting characteristics of these galaxies, such that, galaxies with the reddest colours are ellipticals, and are, generally, metal rich with solar and super-solar values, whereas, galaxies appearing below these ellipticals are S0s, of S- types, which are comparatively bluer due to slow and partial star formation in them, and they tend to be metal poor at sub-solar values. 

Almost all these red sequence galaxies (except the four falling at different loci), with slope of 0.1, represent the underlying stellar populations with varying metallicities. The S- class of objects are observed with slightly bluer colours (statistically) from the passive E counterparts. Interestingly, some ellipticals are seen to overlap and lie almost at the tail of the falling S- objects, which are the low-mass ellipticals with sub-solar metallicities.

The four exceptional galaxies standout from the regular set of points at \textit{bz-yz} $\sim$ 0.15 and \textit{vz-yz} $\sim$ 0.5 due to their excessive blue colours because of the presence of underlying massive, young stars, and not so much due to the change in metallicity. These galaxies could be categorised as the Butcher-Oemler like galaxies. Although the S- galaxies have shown low star forming properties, but such blue star forming elliptical and S0s are indeed surprising. Depending upon the peri-center passage, mass of the halo present in disk galaxies and angle of inclination of the galaxies with respect to the ICM, some disk galaxies have shown evidences of forming stars by molecular gas compression by ram pressure (Fujita \& Nagashima 1999). But these four galaxies are far more bluer than expected to account for this reasoning at their location in the cluster. Moreover, these galaxies are distantly located from their neighbouring galaxies that could influence their star formation, ruling out the possibility of galaxy interactions.

The \textit{vz-yz}, \textit{mz} indices distinguish systems with star formation history of moderate to strong intensity; this plot separates the passive ellipticals from the star forming low-mass objects. There is a small division between the passive ellipticals, S- galaxies and the objects with low star formation, shown by the low values of \textit{mz} ($\sim$ -0.2) and \textit{vz-yz} ($\sim$ 0.5). The two points of S- galaxies at \textit{mz} $\sim$ 0.1 and \textit{vz-yz} $\sim$ 0.5 could probably be the low-mass ellipticals with extremely old age and low metallicity (Rakos et al. 2005b). This colour tool tends to be confusing, especially for such low-mass galaxies with peculiar behaviour.  

Although the \textit{bz-yz}, \textit{uz-vz} colour indices show similar spread of points, like the \textit{bz-yz}, \textit{vz-yz} plot, but with a slightly tighter correlation due to the moderate change in metallicity, pointing to the extra sensitivity to the chemical enrichment of these populations. In a general sense from our studies in other clusters, we have learnt that galaxies with \textit{uz-vz} $\geq$ 0.75, \textit{bz-yz} $\geq$ 0.35 tend to be solar and super-solar metallicities, whereas the sub-solar metallicities lie below that range, however, A1314 shows an exceptional trend. This criteria points to fewer massive galaxies with solar and super-solar metallicities compared to rich clusters. 

The SSP model metallicity tracks from GALEV (Kotulla et al. 2009) are illustrated to verify the obtained inferences from these diagrams. The model tracks, lying linear, show a good match for \textit{bz-yz}, \textit{vz-yz}, \textit{mz} colours, but are found slightly bluer \textit{uz-vz} colours. This could be due the poor treatment of models in the UV wavelength (discussed in \S 4.8) and comparatively higher observational errors in the UV wavelength due to poor atmospheric transmission. 

To sum up, the multicolour diagram displays the ellipticals and the S- galaxies in a linear fashion, such that S- galaxies, comparatively, show slightly bluer colours because of probable underlying star forming populations. Some low-mass, metal-poor ellipticals are also observed with traces of star formation, lying at the tail of S- galaxies. Exceptional cases of galaxies with blue colours are observed to standout, in \textit{bz-yz} and \textit{vz-yz} plot, which are categorised as Butcher-Oemler like galaxies. A distinct separation of passive and low-star forming, low-mass galaxies can be seen in the \textit{mz} and \textit{vz-yz} plot. The \textit{uz-vz} and \textit{bz-yz} plots mark the distinction between the sub-solar and solar (and super-solar) metallicity galaxies, conveying fewer massive, metal-rich galaxies compared to rich clusters. The GALEV SSP model metallicity tracks, on average, show a good match for almost all rest-frame narrowband colour indices, considering that our sample of galaxies are analysed as the superposition of different SSP tracks, while avoiding galaxies with ages less than 3 Gyr and complex star formation scenarios.

\begin{figure}
%\vspace{100pt}
 \centering  % this centres figure in column
  \includegraphics[scale=0.65]{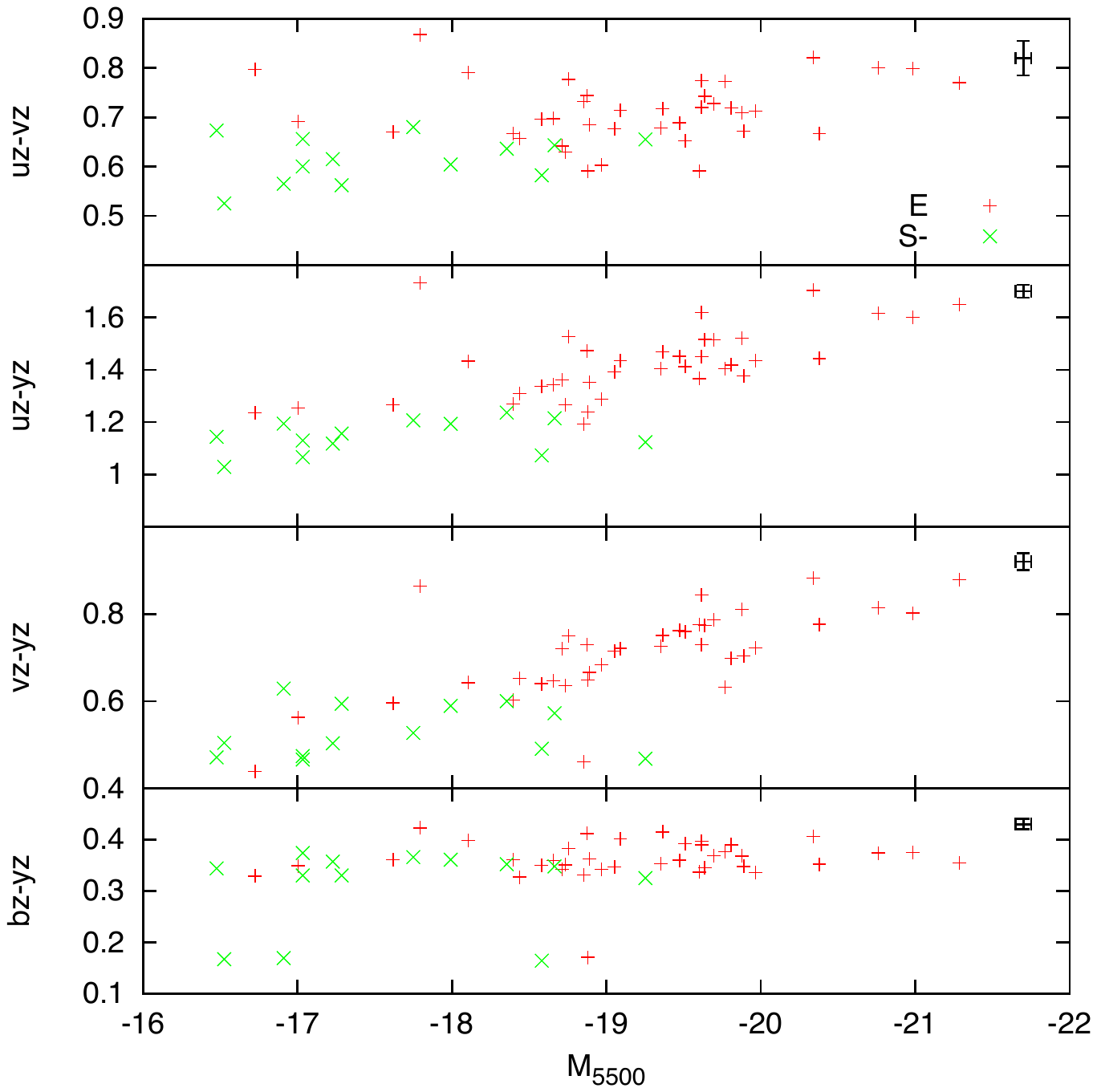}
  \caption{Colour-Magnitude Diagram : The three colour indices are correlated with our \textit{bz} filter at 5500 \AA. A1314 galaxies lie on a linear fashion, with scatter for high- ($\geq$ -19 mag) and low-mass ($\leq$ -19) systems indicating the change in metallicity, age and star formation, under respective colour indices.}
\end{figure}

\subsection{Colour-Magnitude Diagram}

Colour-magnitude Relation (CMR), demonstrated by Faber (1973), Visvanathan \& Sandage (1977), forms a vital tool in understanding the chemical enrichment, as noticed by the change in colour with respect to the galaxy mass. The CMR conveys the ability of galaxy's gravitational potential well to withhold the enriched interstellar medium (ISM) to regulate star formation in these massive systems. In other words, the CMR, especially in the old stellar population, represents the mass-metallicity relation implying more the mass, the greater the ability to collect/retain metallicity.

The CMR for the A1314 cluster sample is shown in Fig. 5 using four colours, \textit{bz-yz} (continuum), \textit{vz-yz} (metallicity indicator), \textit{uz-yz} (near-UV), \textit{uz-vz} (star formation and metallicity index) as a function of $M_{5500}$ luminosity. The ellipticals are, once again, displayed by the red plus symbols and S- are shown in green crosses. The plot shows the points lying in a linear fashion, with small ($\sim$ 0.1) slope for \textit{vz-yz} and \textit{uz-yz} colours. The points show small scatter for all colours below -20 mag, indicating fairly uniform state of formation of these red galaxies in the early Universe. Such low scatter CMR relates to sole metallicity effects, implying the co-evolving red galaxies. 

 Although vaguely visible in our unique modified Str\"omgren colour indices of \textit{vz-yz} and \textit{uz-yz}, a separation between the old, passive, metal-rich and young, star forming, metal-poor populations above and below the $m_{5500}$=-19.2 mag, respectively, similar to what was observed by Rakos et al. (1997; Figs. 3 and 4) for A2317 cluster. 

The \textit{bz-yz} relation with luminosity has been the weakest of all indices, as the points lay almost flat with no clear correlation. However, this could be due to age effecting in response to the mass-metallicity effect, as the \textit{bz-yz} index is more sensitive to the turnoff stars in the old stellar populations. The four exceptional low-mass galaxies with star formation are observed to lie flat, with \textit{bz-yz} $<$ 0.2, hinting to the presence of young stellar population in them.

In conclusion, the CMR presents a not so intense change in metallicity and age in the passively evolving A1314 galaxies. The plot, vaguely, displays two sets of galaxies populations above and below the -19.2 mag mark of old, passive, metal-rich and young, star forming, metal-poor galaxies; both showing almost equal scatter in \textit{uz-vz} and \textit{uz-yz} indices indicating some small secondary star formation events taking place, with small change in metallicity, as displayed by small slope (0.08) for the \textit{vz-yz} colour index. Almost no slope (0.02) for the \textit{bz-yz} colour conveys that most galaxies have been evolving passively with smaller traces of star formation. Some exceptional cases are observed with substantial level of star formation, compared to other galaxies in this cluster, which lie at bottom of the \textit{bz-yz} CMR.

\begin{figure}
%\vspace{100pt}
 \centering  % this centres figure in column
  \includegraphics[scale=0.65]{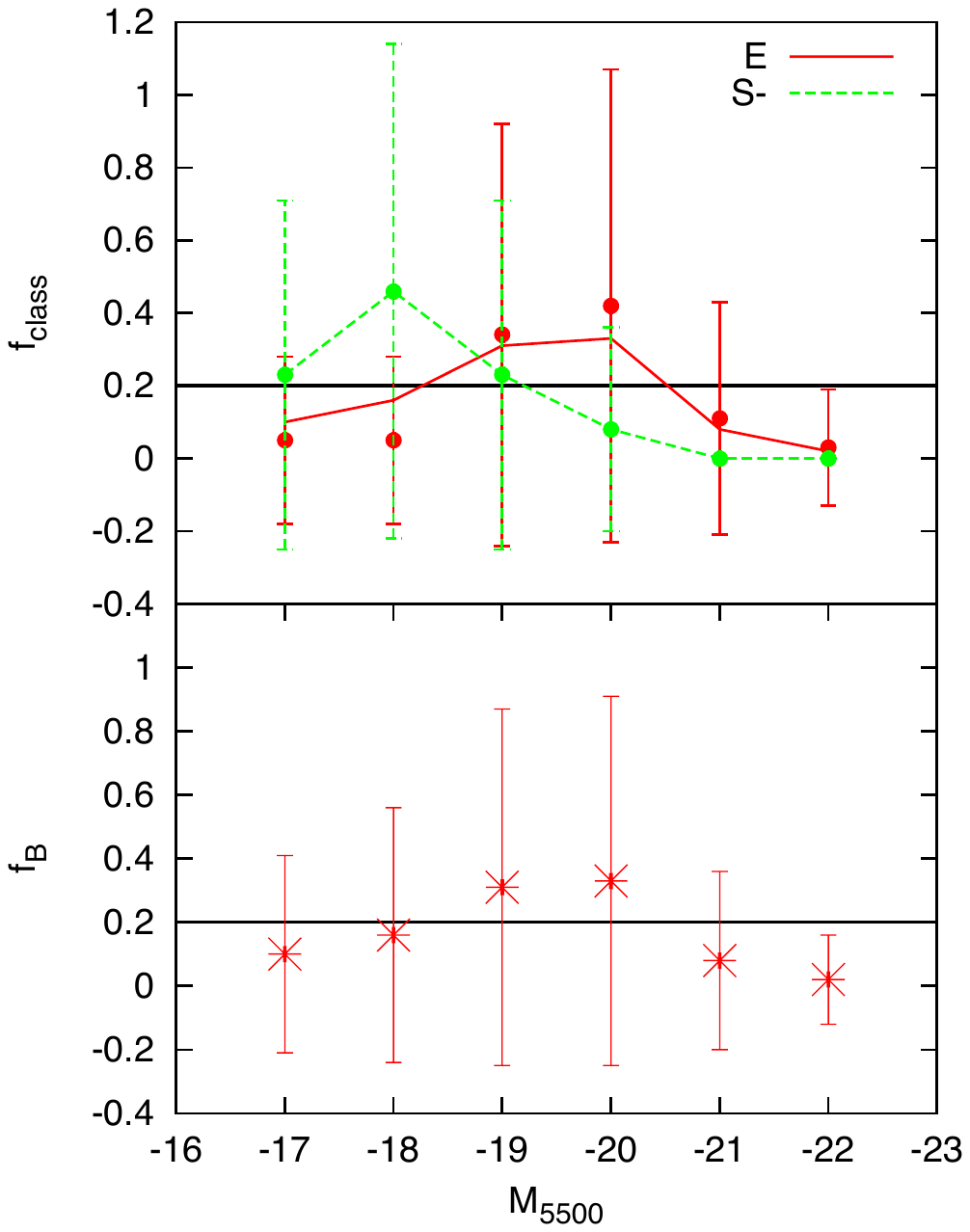}
  \caption{\footnotesize{Upper plot depicts the fraction of spectrophotometric types and lower plot gives the fraction as a function of $M_{5500}$. The horizontal line shows the original work Butcher-Oemler with total $f_B$ for $M_{5500}$ = -20. Overall, the plot indicates the fraction of galaxies with bluer colours, due  to star formation, in red sequence galaxies of A1314. The fall in the number of E class of galaxies is picked by the transition types at low-luminosities that suggests the ongoing star formation in these S- galaxies. The large error bars are due to the small fraction of E and S- types to the total number (upper) and galaxies of both types combined (lower) in the bin size of 1 mag.}}
\end{figure}

\subsection{Butcher-Oemler Effect, or just passive colour evolution?}

Considering the passive state of most galaxies in this cluster, the probability of finding fewer galaxies with bluer colours, would not be a surprise. Butcher \& Oemler (1984) predicted a sharp increase in the blue sequence galaxies at intermediate to high redshift (\textit{z} $\sim$ 1-2) in clusters, and the density of passive red sequence of galaxies would be in abundance at lower redshifts; A1314 is an epitome of such a cluster. The original definition by Butcher \& Oemler for the fraction of red to blue sequence galaxies ($f_B$) is the ratio of galaxies which are 0.2 mag bluer in the broadband (\textit{B-V}) mean E/S0 colours after k-correction to the total number of cluster galaxies. This colour scheme has been transformed by Rakos et al. (2001a), such that, galaxies which are bluer than our rest-frame \textit{bz-yz}=0.22 to the total number galaxies, are referred as the Butcher-Oemler galaxies. 

For A1314, this value is 0.08 with only 4 galaxies of 51 members being bluer than \textit{bz-yz}=0.22. This turns out to be almost similar to our other cluster sample, A2218 (Rakos et al. 2001b), presenting $f_B$ = 0.06. However, the value is still greater than the value predicted for the nearby clusters ($f_B$ = 0.04), but lower compared to our other dense and dynamic cluster samples, like, the young A1185 with $f_B$ = 0.21, A2125 and Coma with 0.15 and 0.1, respectively. This conveys an average to poor performance of star formation activity present in the A1314 galaxies. Fig. 6 displays the Butcher-Oemler properties for A1314 galaxies, the lower panel shows the fraction of blue sequence galaxies ($f_B$) as a function of absolute magnitude (in 1 mag bin size) and the horizontal line gives the total $f_B$ for $M_{5500}$=-20 mag, from the original Butcher-Oemler criteria. The large error bars are because of small fraction of individual galaxy types to the total number (in the upper plot) and similar small fraction of galaxies in 1 mag bin size. They are estimated as the square root of this fraction.

In the current sample, as expected, galaxies with bluer colours have lower luminosities ($<$ -20 mag), and galaxies at higher luminosities ($>$ -20 mag) have passive colours. In the luminosity range -20 $<$ $M_{5500}$ $<$ -21, there is a clear dearth of Starbursts and star forming populations that points to the falling $f_B$ values. This phenomenon is also observed at intermediate redshifts where the blue population divides due to star formation into Spiral and the faint Starburst dwarfs. Often photometric surveys with shallow investigation in the luminosity function have excluded these low-mass dwarfs with blue colours, which are observed in a wide range of luminosities, and present scatter in the $f_B$ plots, either because of the choice of the filter system, response and/or the magnitude cut-off.

The upper part of the Fig. 6 displays the fraction of blue sequence galaxies to the total number of the sample as a function of galaxy class. The plot depicts the same inference of falling numbers of ellipticals on either side of the $M_{5500}$ luminosity range, that is picked up by the transition galaxies at the lower luminosities. Although these transition galaxies are not as blue as the Spirals or Starburst systems, but due to their traces of star formation events (which, perhaps, could be in the stage of their disk formation) they fall into the colour category outlined by Butcher \& Oemler. But the quintessence of this Butcher-Oemler effect is to draw the line between the clusters with active galaxies, like the dwarf Starburst systems, Spirals with heavy star formation, and those which show substantial numbers of Spiral like star formation rates because of the cosmological time (Rakos et al. 2005b). The definition, in general, points and separates galaxy clusters with extreme dynamic state, due to gas rich interactions, mergers, environmental effects, from those of passive and normal clusters, like Coma, A2218 and A1314. 

In conclusion, A1314 galaxy population has been mostly quiescent, with most galaxies showing passive colours, except for four galaxies, which have shown colour criteria outlined by Butcher \& Oemler. These galaxies, even though do not fall in to our colour classification scheme of Starbursts or Spirals, but still, interestingly however, show far bluer colours from the rest of the A1314 population, I term them as Butcher-Oemler like galaxies, which require a deeper probe.

\begin{figure}
%\vspace{100pt}
 \centering  % this centres figure in column
  \includegraphics[scale=0.6]{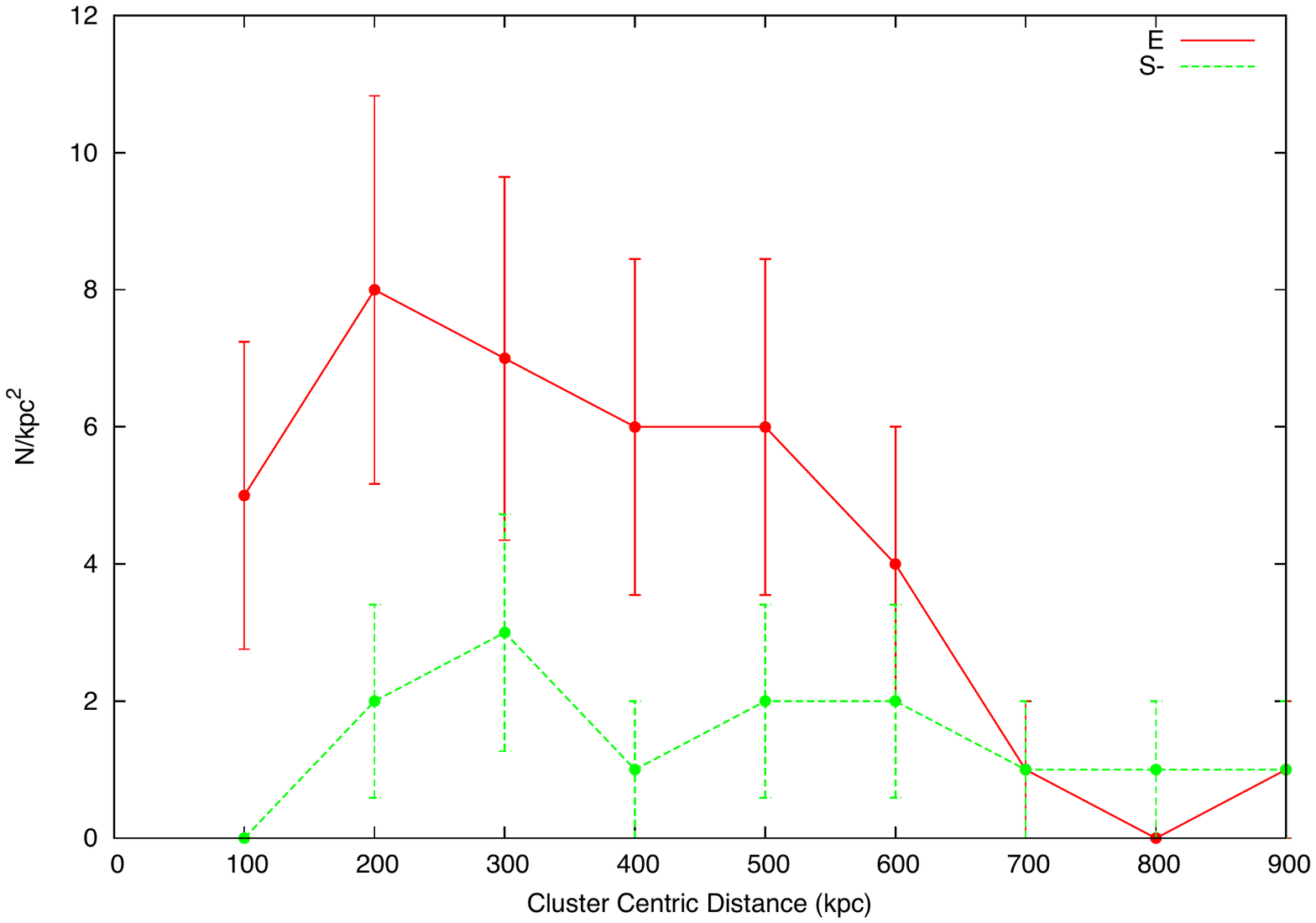}
  \caption{Projected radial galaxy density as a function of cluster centric distance
for each spectrophotometric class. As a general sense, the ellipticals show their crowding near the cluster core, and gradually declining with increasing radial distances. Even with the small number of samples in individual galaxy types (large errors), a general consensus of transition galaxies are observed to avoid core positions, while increasing their numbers outwards.}
\end{figure}

\subsection{Radial Galaxy Density}

The morphology-density relation (Dressler 1984; Smith et al. 1997) suggests 
evolutionary processes maybe strongly related to the environmental effects (Boselli 
\& Gavazzi 2006). Galaxies are known to evolve to differently in different environments 
(clusters, groups and field) and in the presence of other neighbouring galaxies. In fact, 
several authors (Kauffman et al. 2003, 2006; Hogg et al. 2004; Balogh et al. 2004; Baldry 
et al. 2006; Skibba \& Sheth 2009; van der Wel et al. 2010) have shown the influence on 
the structure of galaxies and star formation activity due to tidal interaction with other 
galaxies is much stronger than by the environment. 

Therefore, to study the density of individual galaxy types, though similar to 
the morphology-density relation, despite the small number of samples in individual classes, I 
correlate the the projected radial galaxy density from the cluster centre in Fig. 7. 
The  red dots (with dark line) represent the E-type galaxies and the green dots (with 
dashed line) show the S- galaxies in the 100 kpc$^2$ binned area from cluster centre. 

In line with the general cluster physical scenario, A1314 also follows a similar evolutionary scenario. The inner core ($\sim$100 kpc$^2$) is dominated by the passive, old E 
populations. In the region 100-200 kpc, E galaxies show their maximum numbers, with
some transition galaxies (S-).  At 300 kpc$^2$, the transition galaxy number increase, 
but still being fewer than E types which are observed to gradually decline. At 400 kpc from 
the core, the elliptical numbers are constant, but S- counts are found to drop slightly, 
only to increase at 500 kpc. Beyond this point ($\sim$ 600 kpc), there is steady decline 
in the number of ellipticals, while S- types remain almost constant in their number density.
The falling numbers of ellipticals, near the field or at the edges of the cluster, are 
picked by transition type galaxies showing some star formation activity hinting towards
the blue fraction to the total ratio, as presented in Fig. 6.

This evolutionary picture can be corroborated by the scenarios presented by different 
authors at different wavelengths. The X-ray observations by Forman \& Jones (1999) 
convey the presence of hot ICM spread up to 500 kpc and no cooler gas the centre of 
the cluster, so ruling out the possibility of any star formation activity in this 
region (Bonamente et al. 2002). However, surprisingly, one of the four blue 
Butcher-Oemler like galaxies does show star formation lying within the 500 kpc range, 
perhaps is because the galaxy may not be lying inside the ICM that is spread in 
a highly non-circular manner (Bliton et al 1998). However, these special galaxies do 
require a deeper probe, using Integrated Field spectroscopy (IFS) and Integral Field Unit (IFU) imaging which can resolve in smaller regions and study thoroughly.

\subsection{Age and Metallicity Determination}

While observationally different techniques are able to show similar findings of galaxy properties, their findings diverge by the results obtained using different population synthesis models. When deriving the model ages and metallicities of any stellar population, the initial doubt that comes to mind is the model's reliability to give precise ages and metallicities because of several underlying inadequacies in isochrone and spectral libraries, incorrect and/or inaccurate inputs and treatments of stellar phases at different wavelengths, which collectively are known to affect their output. In addition to those in-built issues, the age-metallicity degeneracy (Worthey 1994) has been a long standing problem that affects models to produce incorrect ages, metallicities (and eventually, the star formation histories). 

Our poorly understood UV properties of the minority population of young stars and, in contrast, the well-studied optical to IR characteristics of the major population (like the Main-Sequence, Red-Giant Branch and Asymptotic-Giant Branch phases of stars) has created a gap in the evolutionary understanding of the stellar systems to raise an astrophysical polemic on the poor mimicking ability of these population synthesis models. This aspect of quickly reddening of colours by the young stars in a majority of old stellar population (the age-metallicity degeneracy), and the effect of this phenomenon on our galaxy age and metallicity estimations, which has been studied by several authors (O'Connell 1980; Gonz\'alez 1993; Worthey 1994; Ferreras et al. 1999; Kuntschner 2000; Trager et al. 2000; Rakos et al. 2005a), has affected the estimates from both methods -- broadband photometry and in the spectroscopy using Lick indices. 

Alternatively, by the use of long-baseline photometry was once thought to be a suitable option to resolve this degeneracy, where in colours from a longer wavelength could separate the effects of age and metallicity, but this attempt has also been challenged because of the following few reasons: 1. At IR wavelengths, the effects of TP-AGB (Thermally Pulsating-Asymptotic-Giant Branch) stars are more prominent, therefore, there is an urgent need to clearly understand the effect of their presence and their contribution in conjunction with stars of other phases. 2. Since there is good contribution by the Main Sequence turn off stars in the UV region, this could be a good age indicator of the underlying old population, but the effects of metallicity and the excess blue colour by the Blue-Horizontal Branch and Blue-Stragglers can result in confusing/wrong age estimates. 3. The small metallicity range incorporated in these models is also a concerning issue, such that, the estimation of ages and metallicities of galaxies at high redshifts, with intricate star formation scenarios and varying chemical enrichment procedures, cannot be reproduced just by using the small metallicity range of globular clusters found inside Milky way and nearby galaxies. A large range of metallicity is a critical requirement for these models.

 Although there has not been a distinct breakthrough in resolving the degeneracy, so far, but as Bruzual (2010) suggests that by properly taking into account the expected variation in the number of stars occupying sparsely populated evolutionary stages, owing to stochastic fluctuations in the Initial Mass Function (IMF), models can well match colours and magnitudes for the entire range of ages and metallicities. Besides that, some alternative tricks to mitigate its effect are being practiced, such as, using the full spectrum fitting technique (Koleva et al. 2008), using surface brightness fluctuation technique (Raimondo et al. 2005; Cantiello et al. 2007), and the use of rest-frame narrowband continuum colours with SSP models supported by the PCA technique (Rakos et al. 1988, 1996, 2001b, 2005a; Rakos \& Schombert 2007; Sreedhar et al. 2012) to name a few. 
 
 \subsection{PCA Supported SSP Models}

Unlike the broadband photometry and spectroscopic lick indices, which have shown significant discrepancies in their age and metallicity estimations, the rest-frame narrowband continuum colours  have displayed fairly smaller counteracting effects of metallicity over age, and vice versa, to get their better measurements. This is because of the narrow filter bandwidth ($\sim$ 200 \AA) rightly placed at the age and metallicity sensitive regions of the SED and the application of the \textit{robust} PCA (Steindling et al. 2001; Rakos et al. 2005a) technique on the SSP models from Schulz et al. (2002), which keep the individual effects as much separate as possible. For extra precaution, populations with young ($<$ 3 Gyr) stars and complex star formation scenarios are avoided because of their extreme blue colours and strong emission lines, which can disrupt our evolutionary analysis.

Moreover, in order to prevent and/or curb such model divergence, our colour indices show sensitivity towards metallicity, dust, 4000 \AA\ break and age of the underlying populations, which are discussed in different articles. In Rakos et al. (1990), they draw a direct, empirical relation to [Fe/H] (between -2.1 $<$ [Fe/H] $<$ 0) using \textit{vz-yz}, \textit{bz-yz} for SSPs and CSPs. In Rakos et al. (2001a), this relation was updated for a much tighter one using the \textit{vz-yz} index with observations of the 41 Milky Way globulars and Fornax dwarf ellipticals (using spectroscopic measurements by Held \& Mould 1994). In Rakos \& Schombert (2004), this updated relation is compared and found to fit well with Schulz et al. (2002) model metallicities in a colour-colour diagram with M87 globulars. Whereas, the \textit{uz-vz} index correlation to the amplitude of spectral [D(4000\AA)] -- a parameter sensitive to the metal content -- is shown in Rakos et al. (2001). The residual \textit{bz-yz} shows a good fit with Milky Way globulars and derived isochrone ages from Salaris \& Weiss (1998) which is illustrated in Rakos et al. (2004). The PCA photometric ages are also compared with Schulz et al. (2002) model ages in Rakos et al. (2005a). In addition, Rakos et al. (2001a) outline a relation of \textit{vz-yz} index with \textit{$Mg_2$} abundance to find that with increasing \textit{vz-yz} colour there is a decrease in \textit{[Mg/Fe]} ratio -- an indirect measure of star formation history. Steindling et al. (2001) point out that since the reddening vector forms at a large enough angle from the effects of age and metallicity, this feature could alleviate the effects of age-dust-metallicity degeneracy (Worthey 1994) by employing PCA technique using rest-frame narrowband colours.

With these photometric benefits, Rakos et al. (2005a) enhanced the Schulz et al. (2002) models using the PCA technique and observed globulars to interpolate between the model ages and metallicities. 

Since the schulz et al./GALEV models were found to be a significant improvement over past works to account for the near-UV and blue colours, which included the recent isochrones from the Padova group that have a stronger blue-HB contribution, (2) use of more recent stellar atmosphere spectra that, again, are refined to better sample the blue part of the SED and (3) are in better agreement with globular cluster \textit{uvby} colours (Rakos et al. 2008). In addition, they offered linearity in the small colour space with the use PC analysis (Steindling et al. 2001) for ages larger than 3 Gyr for a full range of metallicities (Rakos \& Schombert 2005a). In this restricted region, it is possible to apply PC analysis to (1) separate the age and the metallicity of a stellar population, (2) select the most correlated variables, and (3) determine linear combinations of variables for extrapolation. 

However, for cluster galaxy studies, we can isolate galaxies with ongoing star formation, or have large fractions of young stars and dust that can hamper our investigation, by their extreme colours and/or strong emission lines. While most red sequence galaxies are found to contain only small quantity of dust, whose contribution to our colours is only negligible, and is within the order of photometric measurement limits.

Using these models, I have derived the ages and metallicities of A1314 member galaxies, which are verified with the SSP model colour-metallicities tracks for various ages from GALEV (Kotulla et al. 2009) in Fig. 4. In general, the GALEV colour tracks agree well with \textit{vz-yz}, \textit{bz-yz} and \textit{mz} to pass through almost in the middle of the cluster of points, but predicts \textit{uz-vz} slightly bluer than the observations. This disagreement with \textit{uz-vz} could be due to poor knowledge and treatment of Horizontal Branch and Blue Stragglers stellar phases at UV wavelengths, as also illustrated by Carter et al. (2009) who find the divergence in model prediction by 0.05-0.1 mag using broadband filters.

Our comparison of model SSPs of GALEV and Rakos et al. (2005a) of different ages and metallicities presents colour variations of \textit{uz-vz} $\sim$ 0.02, \textit{bz-yz} $\sim$ 0.004, \textit{vz-yz} $\sim$ 0.004 (Sreedhar et al. 2014). This is much smaller than the colour variations reported by Carter et al. (2009) for the broadband \textit{ugriz} and \textit{JHK} filters. Also, our filters, ranging from 3500 \AA\ to 5500 \AA\, are able to determine ages and metallicities with the certainty of 0.5 Gyr and 0.2 dex, respectively, while the combination of broadband filters are able to achieve an inferior ($\Delta$[Fe/H] $\simeq$ 0.18 and $\Delta$Age $\simeq$ 0.25) precision. Whereas, the comparison with spectroscopic Lick IDS studies using SSP spectra from Worthey (1994) with Rakos et al. (2005a) show colours vary as \textit{uz-vz} $\sim$ 0.05, \textit{bz-yz} $\sim$ 0.005, \textit{vz-yz} $\sim$ 0.006 (Sreedhar et al. 2014), while similar discordant results are also observed by Carter et al. (2009) with comparison sample of Trager et al. (2008) who also use isochrones from Worthey (1994). Comparison with latest models, like Flexible Stellar Population Synthesis (FSPS; Conroy et al. 2009), STARLIGHT (Cid Fernandes et al. 2005) are now being tested for agreement with our rest-frame narrowband photometry and will be presented in the subsequent articles.

Even though the considered model (Rakos et al. 2005a using Schulz et al. 2002) is able to reproduce the observed colours well, there is still a small discrepancy in age and metallicity due to the finer colour grids compared to the photometric precision (Sreedhar et al. 2012), that increases with fainter magnitudes. Although these discrepancies are not as severe and significant as compared to the spectroscopic studies, where they show higher numbers of early type galaxies with younger mean ages, in agreement with CDM framework of galaxy formation (Kauffmann et al. 1993; Schiavon 2007). While Bower et al. (1992) found much smaller number of early type galaxies with younger mean ages by the tight CMR. (For more information on the age and metallicity estimation, refer Rakos et al. 2005a and the reference therein.)

In conclusion, by using the PCA supported SSP models for rest-frame narrowband colours, I present the luminosity-weighted mean ages and metallicities of A1314 member galaxies with multiple epochs of star formation; these estimates are the integrated values of the total galaxy luminosity, and not based on the surface brightness. These parameters are found with the model generated mean errors in age of 0.5 Gyr and 0.2 dex in metallicity, which are shown by the error bars at the corner of each plot in Fig. 8. However, a minor effect of the age-metallicity degeneracy still lingers in this technique, based on the precision of the estimated colours to reproduce the age and metallicity parameters of the underlying stellar population, that is, once again, insignificant compared to the effects observed in other evolutionary study techniques.

\begin{table}
\caption{List of A1314 cluster members, the three rest-frame narrowband colours, $M_{5500}$, Concentration Index C, luminosity-weighted mean Age and Metallicity, Galaxy Class, Cluster Centric Distance. Special galaxies with radio dominant, pseudo bulge (denoted by PB), excessive blue Butcher-Oemler (BO) like objects are mentioned in the last column.}
%\small
%\footnotesize
\scriptsize
\centering
\begin{flushright}
\begin{tabular}{llllllllllll}
\hline \\
Sr. & Names & (uz-vz) & (bz-yz) & (vz-yz) & M(5500) & C &  Age   &   Z    & Cls & Dist. & Notes \\ 
 &  &  &  &  & mag & \multicolumn{1}{l}{} & Gyr & dex &  & kpc &  \\ 
 &  &  &  &  &  & \multicolumn{1}{l}{} &  &  &  &  &  \\ 
 \hline \\
1 &  J113545.76+485102.5 & 0.642 & 0.342 & 0.720 & -18.71 & 2.75 & 5.12 & 0.12 & E & 566.7 &  \\ 
2 &  J113545.83+490903.1 & 0.743 & 0.345 & 0.773 & -19.64 & 3.03 & 7.35 & 0.12 & E & 450.6 &  \\ 
3 &  J113543.93+485058.2 & 0.562 & 0.330 & 0.594 & -17.29 & 2.47 & 9.08 & -0.48 & S- & 561.5 &  \\ 
4 &  J113543.96+490215.4 & 0.773 & 0.377 & 0.632 & -19.77 & 2.84 & 6.27 & 0.16 & E & 350.3 & PB \\ 
5 &  J113543.71+490959.8 & 0.718 & 0.415 & 0.751 & -19.37 & 2.81 & 10.79 & 0.04 & E & 461.4 & PB\\  
6 &  J113541.99+490440.6 & 0.677 & 0.347 & 0.715 & -19.05 & 2.76 & 12.56 & -0.26 & E & 350.3 &  \\ 
7 &  J113539.58+485527.7 & 0.591 & 0.337 & 0.775 & -19.60 & 3.04 & 13.00 & -0.30 & E & 417.0 &  \\ 
8 &  J113540.03+490552.6 & 0.689 & 0.360 & 0.762 & -19.47 & 3.04 & 5.00 & 0.22 & E & 354.0 &  \\ 
9 &  J113538.40+485753.4 & 0.603 & 0.342 & 0.684 & -18.97 & 2.93 & 13.11 & -0.38 & E & 358.1 &  \\ 
10 &  J113526.23+490513.3 & 0.672 & 0.348 & 0.704 & -19.89 & 3.16 & 10.12 & -0.18 & E & 263.3 &  \\ 
11 &  J113507.93+490340.0 & 0.797 & 0.329 & 0.439 & -16.73 & 2.53 & 11.99 & -0.54 & E & 131.9 &  \\ 
12 &  J113508.61+491718.4 & 0.709 & 0.368 & 0.811 & -19.88 & 2.88 & 10.86 & 0.02 & E & 595.9 &  \\ 
13 &  J113506.35+490327.8 & 0.667 & 0.361 & 0.602 & -18.40 & 3.13 & 9.74 & -0.28 & E & 119.5 &  \\ 
14 &  J113502.84+490731.7 & 0.604 & 0.361 & 0.589 & -17.99 & 2.11 & 7.04 & -0.20 & S- & 221.5 & \tiny{Excess SDSS} \\ 
%  &                      &       &       &       &        &      &      &       &    &       & \tiny{\textit{u-r} colours}\\ 
   &                      &       &       &       &        &      &      &       &    &       & \scriptsize{\textit{u-r} colours}\\ 
15 &  J113459.64+485919.3 & 0.657 & 0.327 & 0.652 & -18.44 & 2.35 & 6.91 & -0.08 & E & 133.0 &  \\ 
16 &  J113459.87+490452.9 & 0.719 & 0.390 & 0.699 & -19.81 & 3.10 & 6.76 & 0.16 & E & 122.1 & IC712 \\ 
17 &  J113458.11+490545.5 & 0.600 & 0.374 & 0.466 & -17.03 & 1.17 & 9.94 & -0.58 & S- & 146.7 &  \\ 
18 &  J113457.53+491235.6 & 0.565 & 0.169 & 0.629 & -16.91 & 2.80 & 8.73 & -0.82 & S- & 403.7 & BO  \\ 
19 &  J113453.54+490351.1 & 0.691 & 0.349 & 0.563 & -17.00 & 2.07 & 5.17 & -0.02 & E & 67.4 &  \\ 
20 &  J113452.42+491403.9 & 0.685 & 0.362 & 0.666 & -18.89 & 2.65 & 4.92 & 0.12 & E & 457.5 &  \\ 
21 &  J113450.38+490303.1 & 0.629 & 0.351 & 0.636 & -18.73 & 2.57 & 13.10 & -0.42 & E & 31.3 &  \\ 
22 &  J113449.29+490439.4 & 0.770 & 0.355 & 0.879 & -21.29 & 3.51 & 9.15 & 0.18 & E & 92.2 &  \\ 
23 &  J113449.01+490452.9 & 0.777 & 0.383 & 0.750 & -18.76 & 1.90 & 10.79 & 0.04 & E & 100.8 &  \\ 
24 &  J113448.65+491430.1 & 0.680 & 0.366 & 0.527 & -17.75 & 2.09 & 6.54 & -0.10 & S- & 473.9 &  \\ 
25 &  J113446.55+485721.9 & 0.801 & 0.374 & 0.814 & -20.76 & 3.22 & 12.39 & 0.00 & E & 191.0 & IC711 \\ 
26 &  J113442.97+490527.0 & 0.714 & 0.402 & 0.721 & -19.09 & 3.00 & 11.15 & -0.06 & E & 129.0 &  \\ 
27 &  J113442.97+490608.3 & 0.673 & 0.344 & 0.471 & -16.48 & 2.05 & 3.70 & -0.10 & S- & 154.6 &  \\ 
28 &  J113442.62+490313.0 & 0.670 & 0.361 & 0.596 & -17.62 & 2.37 & 12.64 & -0.36 & E & 54.9 &  \\ 
29 &  J113435.54+491320.9 & 0.791 & 0.399 & 0.642 & -18.11 & 2.46 & 6.91 & 0.16 & E & 437.9 &  \\ 
30 &  J113433.94+490515.6 & 0.744 & 0.412 & 0.729 & -18.87 & 2.95 & 4.61 & 0.32 & E & .5 & \\ 
31 &  J113432.97+490605.5 & 0.652 & 0.392 & 0.760 & -19.51 & 2.98 & 7.40 & 0.12 & E & 179.9 &  \\ 
32 &  J113431.00+490739.9 & 0.720 & 0.397 & 0.730 & -19.62 & 2.50 & 13.26 & -0.16 & E & 238.6 &  \\ 
33 &  J113427.99+485748.2 & 0.678 & 0.353 & 0.726 & -19.35 & 3.17 & 4.99 & 0.16 & E & 219.6 &  \\ 
34 &  J113421.48+490733.5 & 0.636 & 0.352 & 0.600 & -18.36 & 2.22 & 12.45 & -0.42 & S- & 269.8 &  \\ 
35 &  J113416.20+490014.3 & 0.728 & 0.369 & 0.787 & -19.70 & 3.13 & 10.83 & 0.02 & E & 224.1 &  \\ 
36 &  J113416.27+490202.1 & 0.696 & 0.350 & 0.640 & -18.58 & 2.62 & 5.04 & 0.08 & E & 209.2 &  \\ 
37 &  J113414.53+490235.4 & 0.799 & 0.375 & 0.802 & -20.98 & 3.34 & 12.52 & -0.01 & E & 220.1 & IC709 \\ 
38 &  J113411.80+490303.1 & 0.656 & 0.330 & 0.474 & -17.03 & 2.06 & 9.69 & -0.58 & S- & 238.8 &  \\ 
39 &  J113408.87+491515.9 & 0.643 & 0.348 & 0.572 & -18.66 & 2.61 & 7.34 & -0.22 & S- & 564.8 &  \\ 
40 &  J113403.86+490444.3 & 0.615 & 0.357 & 0.503 & -17.23 & 2.12 & 4.65 & -0.16 & S- & 302.1 &  \\ 
41 &  J113403.63+490159.4 & 0.697 & 0.359 & 0.646 & -18.66 & 2.57 & 5.23 & 0.10 & E & 289.2 &  \\ 
42 &  J113402.15+490907.7 & 0.774 & 0.390 & 0.844 & -19.62 & 3.58 & 12.80 & 0.02 & E & 399.5 &  \\ 
43 &  J113400.24+490349.8 & 0.868 & 0.423 & 0.864 & -17.80 & 3.34 & 7.51 & 0.42 & E & 315.6 & IC708 \\ 
44 &  J113343.62+485602.7 & 0.713 & 0.336 & 0.722 & -19.97 & 2.86 & 10.88 & -0.16 & E & 483.3 & PB \\ 
45 &  J113323.02+490217.0 & 0.732 & 0.331 & 0.461 & -18.86 & 2.53 & 8.23 & -0.44 & E & 549.1 & E7/ \\ 
   &                      &       &       &       &        &      &      &       &   &       & \scriptsize{Edge-on PB} \\ 
% &  &  &  &  &  &  &  &  &  &  &  \\ 
\multicolumn{1}{l}{} & \multicolumn{1}{l}{} & \multicolumn{1}{l}{} & \multicolumn{1}{l}{} & \multicolumn{1}{l}{} & \multicolumn{1}{l}{} & \multicolumn{1}{l}{} & \multicolumn{1}{l}{} & \multicolumn{1}{l}{} & \multicolumn{1}{l}{} & \multicolumn{1}{l}{} & \multicolumn{1}{l}{} \\ 
\hline \\
\end{tabular}
\end{flushright}
\end{table}
\newpage

\begin{table}
%\small
%\footnotesize
\scriptsize
\centering
\begin{flushright}
\begin{tabular}{llllllllllll}
\hline \\
Sr. & Names & (uz-vz) & (bz-yz) & (vz-yz) & M(5500) & C &  Age   &   Z    & Cls & Dist. & Notes \\ 
 &  &  &  &  & mag & \multicolumn{1}{l}{} & Gyr & dex &  & kpc &  \\ 
 &  &  &  &  &  & \multicolumn{1}{l}{} &  &  &  &  &  \\ 
 \hline 
%41 &  J113403.63+490159.4 & 0.697 & 0.359 & 0.646 & -18.66 & 2.57 & 5.23 & 0.10 & E & 289.2 &  \\ 
%42 &  J113402.15+490907.7 & 0.774 & 0.390 & 0.844 & -19.62 & 3.58 & 12.80 & 0.02 & E & 399.5 &  \\ 
%43 &  J113400.24+490349.8 & 0.868 & 0.423 & 0.864 & -17.80 & 3.34 & 7.51 & 0.42 & E & 315.6 & IC708 \\ 
%44 &  J113343.62+485602.7 & 0.713 & 0.336 & 0.722 & -19.97 & 2.86 & 10.88 & -0.16 & E & 483.3 & PB \\ 
%45 &  J113323.02+490217.0 & 0.732 & 0.331 & 0.461 & -18.86 & 2.53 & 8.23 & -0.44 & E & 549.1 & E7/ \\ 
%   &                      &       &       &       &        &      &      &       &   &       & \scriptsize{Edge-on PB} \\ 
46 &  J113322.19+485739.1 & 0.591 & 0.171 & 0.649 & -18.88 & 3.00 & 5.62 & -0.42 & E & 583.2 & BO  \\ 
47 &  J113313.49+490606.9 & 0.821 & 0.406 & 0.882 & -20.34 & 3.18 & 11.45 & 0.20 & E & 627.2 &  \\ 
48 &  J113304.77+490509.4 & 0.525 & 0.167 & 0.504 & -16.53 & 2.37 & 8.75 & -1.16 & S- & 674.5 & BO \\ 
49 &  J113251.84+485333.7 & 0.582 & 0.164 & 0.491 & -18.58 & 2.36 & 7.89 & -1.08 & S- & 822.3 & BO  \\ 
50 &  J113249.79+490601.5 & 0.655 & 0.325 & 0.468 & -19.25 & 2.28 & 3.30 & -0.06 & S- & 774.6 &  \\ 
51 &  J113234.81+490634.7 & 0.667 & 0.352 & 0.776 & -20.38 &  & 5.22 & 0.20 & E & 872.5 &  \\ 
% &  &  &  &  &  &  &  &  &  &  &  \\ 
\multicolumn{1}{l}{} & \multicolumn{1}{l}{} & \multicolumn{1}{l}{} & \multicolumn{1}{l}{} & \multicolumn{1}{l}{} & \multicolumn{1}{l}{} & \multicolumn{1}{l}{} & \multicolumn{1}{l}{} & \multicolumn{1}{l}{} & \multicolumn{1}{l}{} & \multicolumn{1}{l}{} & \multicolumn{1}{l}{} \\ 
\hline \\
\end{tabular}
\label{}
\scriptsize
      \hfill\parbox[ht]{16cm}{Note: RA and declination of individual galaxies are deliberately omitted because the SDSS galaxy names already include their respective \\ coordinates. For e.g., the IC708 galaxy with the SDSS name J113400.24+490349.8 gives RA 11:34:00.24 and declination +49:03:49.8.}
\end{flushright}
\end{table}
%\end{landscape}

\begin{figure}
%\vspace{100pt}
 \centering  % this centres figure in column
  \includegraphics[scale=0.65]{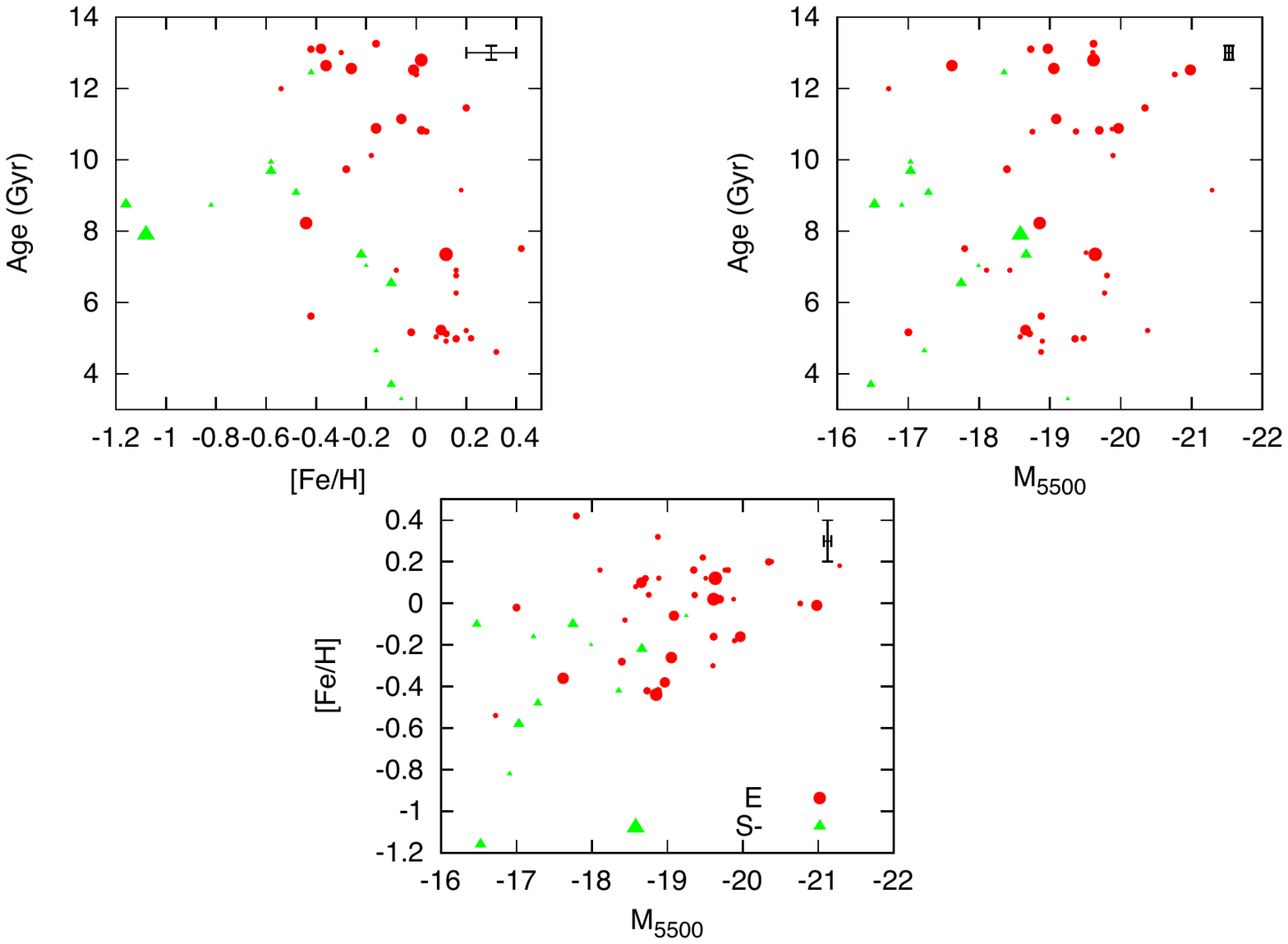}
  \caption{Age, metallicity, mass (luminosity) plot for A1314 galaxies. The red dots
are the ellipticals and the green triangles are the transition galaxies. The differing 
sizes of the data points show the true dispersions in the averages of the SSP ages and metallicities, which represent the star formation history, such that, the bigger the data size, longer the star formation history. Even in all red sequence galaxy cluster member, there is bimodality visible. The plot depicts older galaxies are massive and younger galaxies are low-mass, and there is a scattered and crude representation of mass-metallicity relation.}
\end{figure}

\subsection{Age and Metallicity Analysis}

The estimated ages and metallicities of 51 A1314 member galaxies are listed in the Table 2 and illustrated in Fig. 8, along with luminosity $M_{5500}$, as mass. The estimated errors, generated from the method, are displayed at the corner of each diagram. The red dots are the ellipticals and the green triangles represent the S- transition galaxies. The differing sizes of the data points in the plots show the true dispersions in the averages of the SSP ages and metallicities, and are not uncertainties. Since these true dispersions depict the standard deviation of the averaged SSP ages and metallicities, they would represent the star formation history, in such a way, that the higher the dispersion due to longer star formation history, the bigger the size of the data point.

In the age-metallicity plot, even though the cluster hosts almost all red sequence galaxies, there is still a vague (but in accordance to other cluster studies) bimodal population visible; the first set of galaxies are very old ($>$ 10 Gyr) for which the metallicities ranging from solar to -0.4 dex, and, the other population being younger than 10 Gyr showing a large chemical evolution, ranging from 0.2 to -1.2 dex. The very old population (with small span in metallicity) must have formed its majority population very early in the Universe with a short instant burst of star formation. The narrow gap between the two sets of population is scattered with a few passive, low-mass galaxies with low star formation history; these galaxies are slowly ($\sim$ 1 Gyr) evolving in metallicity (from solar to sub-solar metallicity). Beyond this dormant phase, a gradual rise in the numbers of young population is noticed with a fair chemical enrichment, ranging from -0.6 to 0.4 dex. 

A few transition types of intermediate age ($\sim$ 8 Gyr) are also found, which lie astray and are extremely metal poor, with substantial star formation history. These galaxies are expected to lie beyond the realm of hot ICM, at distances of more than 500 kpc from core. In a crude sense, the transition galaxies show their chemical evolution from solar metallicity during their young ($<$ 5 Gyr) ages to intermediate age (7-10 Gyr) with extremely metal poor values, with substantial star formation history, to finally end up as passive galaxies with similar trends like the old ellipticals of 10-12 Gyr with solar and sub-solar metallicities. The presence of young ellipticals with solar metallicities represent the in-falling galaxies, which had their secondary bursts of star formation, are now in the inner 500 kpc region tidally stripped by the interaction with the ICM, with their star formation quenched to retain high metallicities, similar to their older counterparts.

The age-luminosity plot, once again, shows the bimodal population of galaxies: those which are very old are high-mass and, the ones, which are young with some secondary star formation activity are low-mass. The in-between region of the intermediate age (7-10 Gyr) is scattered with a few transition galaxies of moderate star formation history, but slightly higher than the ellipticals of the same age. The low-mass ($<$ -20 mag), young ($<$ 8 Gyr) ellipticals resemble transition galaxies in their star formation property, with similar point size. Their loci in this plot are usually occupied by S0-type galaxies in other poor clusters (for e.g., A779 of our sample) and in rich clusters by spirals (in A2218, A2125, A1185). This property of low-mass ellipticals in A1314 is, indeed, strange and they might be the blue sequence of ellipticals or pseudo bulges; a deeper analysis using multi-wavelength spectrophotometric surveys is recommended.

The scatter of points of the very old ($\geq$ 12 Gyr) ellipticals, with bigger point size, convey, that these galaxies could have either formed with a short burst of intense star formation and later evolved by accumulating mass of other similarly aged galaxies in the vicinity, by minor mergers, to have a luminosity-weighted mean age of $\geq$ 12 Gyr, or else, the entire mass itself of the galaxy could have formed by one epoch of star formation, lasting for 12 Gyr -- similar to the prediction by the monolithic collapse structure formation scenario (Eggen et al. 1962; Larson 1974; Arimoto \& Yoshii 1987). Whereas, some other galaxies, with smaller point size, under the same age range, suggest that these massive galaxies are formed with a short burst of star formation at very early in the Universe, and did not accumulate mass over time, otherwise, these points would have had much younger mean ages. 

Nevertheless, these old galaxies are found to have varying mass, from -16 to -21 mag. Similarly, the points presented by the young, low-mass galaxies suggest these galaxies, once again, could have had one or more epochs of star formation, lasting either 8 Gyr or shorter, to combine with underlying old stellar population to give a luminosity-weighted mean age of $\leq$ 8 Gyr. Interestingly, in A1314, the presence of hot ICM in the inner regions, gives a way to understand the situation better, such that, the galaxies which reside in the pool of ICM would not have formed stars; so, whatever star formation events the galaxy witnessed, must have been before its arrival to central inner regions. Of course, this idea would also depend upon the period and axis of its rotation of the galaxy around the core. However, the base of the initial star formation in these central galaxies would now be old stellar populations, which, when assembled into the cluster, quenched their star formation by encountering the ICM to be left for passive evolution by gathering mass via merging of similarly aged stellar populations, resembling the classical bulges.

Concurrently, there are also few S- or transition galaxies in the low-mass population (the big triangles), which are still young because of their long epochs of star formation history, these galaxies should be residing in regions beyond 500 kpc (discussed in greater detail in the next section). These galaxies are the successors of the massive galaxies seen in the centre in the ICM pool.

However, understanding the trend of age with stellar mass is more difficult to extract from our data because the integrated age of the underlying population can be reached by many avenues -- either by simultaneous formation to represent our age measurement to that time epoch or if galaxies form their initial stars over a range of time, (but all starting and stopping at the same epoch), then the age would be the median time of the initial star formation epoch. 

This conundrum relates to the debate of the formation time of the stellar population versus the assembly time of those populations into their parent galaxy (De Lucia et al. 2006). But recently, a variety of hierarchical models have been considered in which stars form in smaller units, then merge into their final systems. Nevertheless, the trend of old ($\geq$ 10 Gyr), massive galaxies degrades, although slightly, to the younger ($\leq$ 8 Gyr), low-mass counterparts is the "downsizing' scenario of galaxy formation (Bundy et al. 2005; Mateus et al. 2006).

The metallicity-luminosity plot depicts the galaxy mass-metallicity relation (Nelan et al. 2005; Kelson et al. 2006) where galaxies tend to grow metal rich with increasing mass, though not linearly because of the observed scatter in the CMR (Fig. 5). As there is no effect of age in the CMR, the parameter does not play a major role in this mass-metallicity relation. Because of this, there are many young galaxies, which, in fact, show higher metallicities than the old galaxies. Nevertheless, the plot depicts, with some scatter, that massive galaxies are found to have increasing metallicity.

\begin{figure}
%\vspace{100pt}
 \centering  % this centres figure in column
  \includegraphics[scale=0.65]{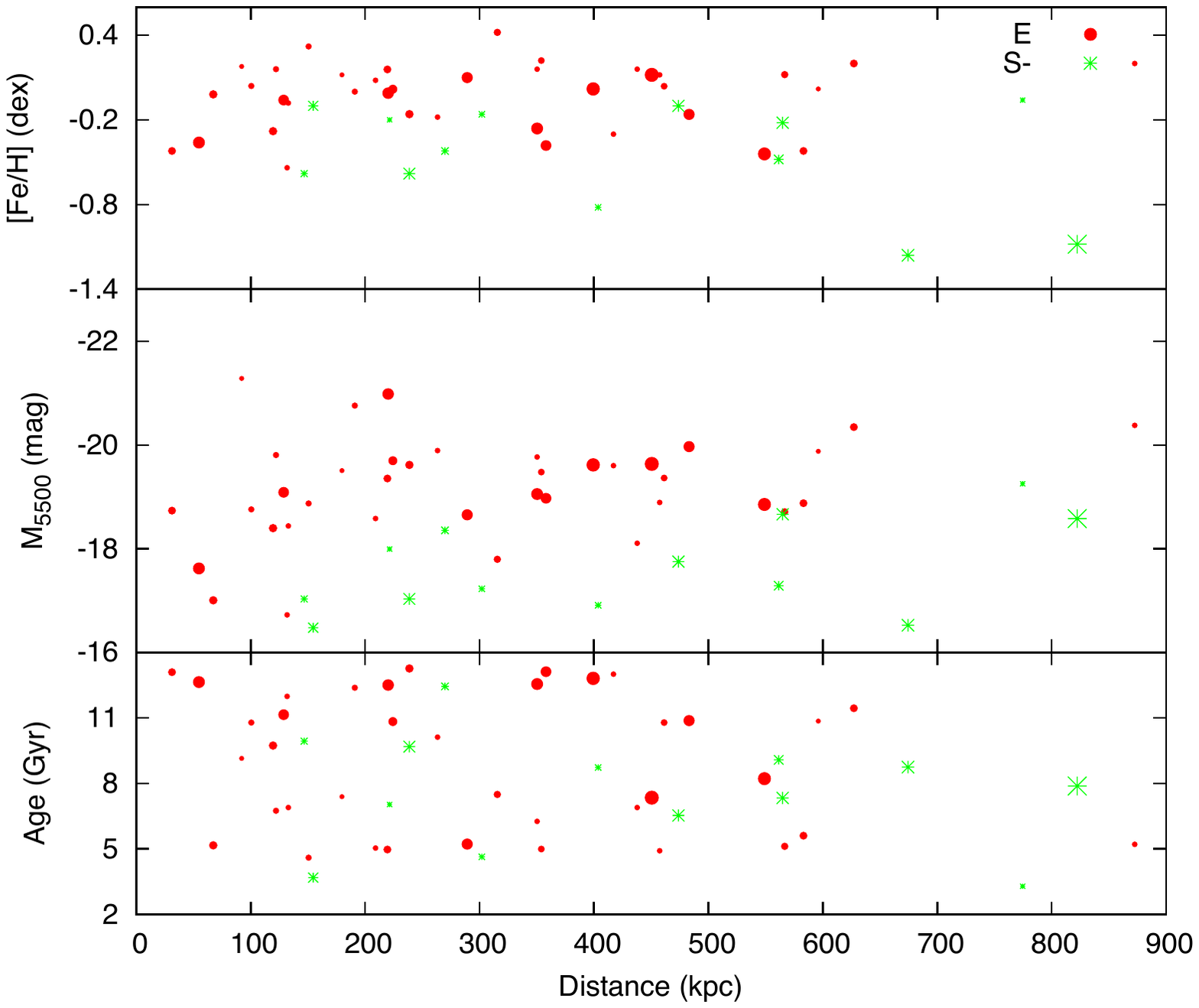}
  \caption{Age, mass and metallicity distribution across the cluster radial distance as
a function of spectrophotometric class and star formation history. The bottom most plot shows the inner cluster regions are filled with old ellipticals, whereas the star forming, intermediate age transition galaxies are found beyond 500 kpc from the cluster. The middle plot depicts the massive ellipticals are present in the core, whereas, the low-mass transition systems gradually increase in numbers at distances $>$ 400 kpc. Solar and super-solar metallicity galaxies are found to crowd the inner regions, whereas the metal poor transition galaxies are found at distances $>$ 400 kpc.}
\end{figure}

\subsection{Age, Metallicity and Mass Correlation with Distance}

To have a clear understanding of gradients in age, metallicity and mass across the cluster, I correlate these terms with cluster centric distance as a function of star formation history, which is displayed in Fig. 9. These correlations supports the hints gathered from earlier plots, especially from the age, metallicity and mass correlation (Fig. 8), and the inferences obtained from X-ray and radio observations. 

Starting from the bottom most plot, the inner cluster regions ($\sim$ 300-400 kpc) are filled with many ellipticals, as expected, and fewer S-/transition galaxies. These galaxies had ceased their star formation long ago and are now evolving passively. The number of transition galaxies increases beyond the 400 kpc mark, where the ellipticals and S- types show moderate star forming history. This can also be perceived by the middle plot of mass distribution across the radial distance, where the number density of ellipticals gradually increases from the core to reach a maximum at 300 kpc$^2$, that holds the maximum mass, including a few transition galaxies. 

On a careful inspection of all three plots, one can see that there is a presence of at least one (or more) big red dot in every 100 kpc$^2$ area conveying that there is at least one (up to a distance 
of 600 kpc from cluster centre) big elliptical galaxy in every 100 kpc$^2$ that while drifting towards the cluster centre is accumulating mass by mergers with low mass ellipticals (found in abundance in the 
inner 300 kpc$^2$ region of the cluster). And, because of this accumulation, the big elliptical reflects an extended star formation history, hence the big red dot. In the 100-200 kpc$^2$ region, the mass merging creates ellipticals with luminosity-weighted mean metallicity of 0.2 to -0.6 dex. On the other hand, the transition galaxies are distributed randomly across the cluster, which show increasing, and perhaps ongoing, star formation activity with increasing distance from the cluster centre.

The top most plot of metallicity gradients shows a zig-zag trend to interpret that these low-mass, metal rich (solar and super-solar metallicities) ellipticals which when fall inside the cluster core, encounter the ICM, get tidally stripped off their gas to be quenched of their star formation. These galaxies evolve slowly, with steady galaxy interaction, perhaps with dry, minor mergers  -- an instance where red and dead ellipticals are formed. These types of galaxies are observed within the 300-500 kpc$^2$ regions with most of their gas has been shock-heated at the virial temperature off the clusters and are observed in X-rays.

The radio prominent galaxies, IC 712, IC 711, IC 709, IC 708 are found to lie at 100, 191, 220 and 315 kpc distance from the centre, respectively, which do not show any optically interesting features. IC 708 is a low-mass elliptical of age of 7.5 Gyr and metallicity of 0.42 dex, probably located at the edges of the variably  spread ICM, had formed stars and regulated chemical enrichment before impending the interaction with the ICM. The rest three galaxies are old ellipticals with passive evolution located much closer to the cluster core. On other hand, the close proximity to the cluster's gravitational potential and the presence of ICM, creates a low to moderate star formation history for these three galaxies. In most radio studies, IC 712 is considered as the galaxy nearest to the cluster centre, which is not quite correct, as optically detected J113449.29+490439.4 of -21 mag and the mean age of 9.2 Gyr is the massive galaxy in A1314, and is closest to the core.

The four Butcher-Oemler like galaxies are all young, low-mass galaxies with solar to super-solar metallicities. All four galaxies are spread out beyond 400 kpc, the one (J113251.84+485333.7) that lies the farthest (822 kpc) is the transition galaxy, that shows the most bluest colours with substantially high star formation history, compared to others. These Butcher-Oemler like galaxies do show capabilities of being blue sequence E/S0s due to their Butcher-Oemler colours, and their fuzzy structures with halo around them. 

Besides these, there are also galaxies, which show similarities to pseudo bulges, such as for example, J113543.96+490215.4, J113343.62+485602.7, J113543.71+490959.8 and, maybe, J113323.02 +490217.0 as they resemble galaxies with bar and/or disks, with ages between 6.3-10.8 Gyr showing moderate to low star formation history. All these probable pseudo bulges resides at far out reaches from the variably distributed ICM. This has been identified by comparing the observed \textit{y}-band image with X-ray exposure of Jones \& Forman (1999). 

Some misclassification of their morphology could be possible for galaxies with disks, which are mostly located beyond 500 kpc. To confirm and further understand the properties of these read and dead object in the core and pseudo bulges at the edges of the cluster, more in-depth observations are required using multi-wavelength, high spatial resolution photometric surveys.

In conclusion, I find using our rest-frame narrowband colours and PCA supported SSP models of Rakos et al. (2005a), that galaxies in this moderately dense A1314 cluster had initially formed stars early in the Universe, then assembled in the cluster and while encountering the ICM, their star formation quenched, turned passive to gather mass by dry, minor mergers with low mass galaxies at later epochs. These A1314 member galaxies can be perceived in three different sections on the basis of their cluster centric distance: first, the very old, massive systems (which could be red and dead objects) located near the core ($\sim$ 200 kpc) with metallicities about solar values are evolving passively, the second, galaxies between 200-500 kpc are the recently fallen ellipticals and transition types, which are (or are in the process) tidally gas stripped, showing low to moderate star formation history and low to solar metallicities, and the third section consisting of substantial star forming galaxies, which are beyond the realm of the ICM ($>$ 500 kpc) are slowly drifting in-wards, with a range of age (8.7-11.5 Gyr), metallicity (0.2 - -1.2 dex) and mass (-16.5 - -20.3 mag), showing signs of bars and a disks to resemble pseudo bulges.

\section{Conclusions}

The radio bright cluster of A1314 is famous for presenting one of the longest head-tail radio galaxy with soft excess low X-ray emission (at 2-6 keV band $L_x$ $\le$ 7 $\times$ $10^{44}$ erg $s^{-1}$); the cluster has a richness class III and Bautz-Morgan class 0 at \textit{z}=0.034. In this study, the cluster is analysed primarily using the rest-frame narrowband colours for its passive evolutionary nature of hosting almost all members (identified with the \textit{mz} index) as the red sequence galaxies -- classified using the PCA technique and also recognised using the SDSS colour correlation of \textit{u-r} and the concentration index. As a primary objective to understand the formation of red sequence galaxies under the influence of internal and environmental effects, the luminosity-weighted mean ages and metallicities of member galaxies are derived using the PCA supported SSP models of Schulz et al. (2002) developed by Rakos et al. (2005a). By correlating various parameters, measured and derived, I draw the following inferences: 

\begin{enumerate}
\item The multicolour diagram displays a linear arrangement of ellipticals and the S- (transition) types with traces of star formation in their underlying populations. Some exceptional cases with excessive blue colours standout from the normal bunch of galaxies in \textit{bz-yz} and \textit{vz-yz} colour plots. The \textit{mz} and \textit{vz-yz} plot shows separation of passive and low-star forming galaxies, whereas, the \textit{uz-vz} and \textit{bz-yz} plot points to fewer massive galaxies with solar and super-solar metallicities. The GALEV SSP model metallicity tracks are found to agree well with almost all colour modified Str\"omgren colour indices, except for the \textit{uz-vz} index which are predicted bluer than observed. This could be because of poor treatment of blue-horizontal branch and blue straggler stellar phases in the UV wavelength and comparatively larger observational errors due to poor UV atmospheric transmissions.

\item The CMR presents, although vaguely, a separation of populations of passive, old, metal-rich and star forming, young, metal-poor types on either side of the -19.2 mag. Also, the shallow slopes and scatter of points in \textit{uz-vz} and \textit{uz-yz} colour convey a small change in metallicity and age due to traces of secondary star formation. Exceptional cases are observed with moderate to substantial levels of star formation, which, owing to this property, lie flat at bottom of the \textit{bz-yz} CMR.

\item The Butcher-Oemler classification scheme finds most galaxies as quiescent with passive colours. Only four galaxies are found to have the colour criteria outlined by Butcher and Oemler. Although unlike Spirals and Starbursts, these galaxies are substantially blue which I term them as Butcher-Oemler like galaxies.

\item The radial galaxy density profile illustrates a general picture (like, our other studied clusters A779, A539, A400) that the inner cluster regions are occupied by the passive ellipticals, which  gradually decline in their numbers with increasing distances. The transition galaxies show a slow increase in their number density extending radially outwards.

\item In the age, metallicity, mass plot, a bimodal distribution is observed for galaxies of old ($\geq$ 10 Gyr) with solar to -0.4 dex range in metallicity and young ($\leq$ 8 Gyr) with a wide range of chemical evolution (0.2 to -1.2 dex). 

\item The in-between region of old and young population is scattered by a few low-mass galaxies with low star formation history. The small range in metallicity of the very old galaxy convey either to have formed entirely via short intense burst of star formation or would have accumulated mass of similarly aged stellar populations over the cosmic time. 

\item The very old passive galaxies are found to be massive compared to the young ones, presenting the downsizing behaviour of galaxy mass. A scattered mass-metallicity relation can be observed, since age does not play a vital role, to find that the brightest galaxies are metal-rich.

\item The transition galaxies evolve in their chemical enrichment with solar metallicity at 5 Gyr to turn  metal poor at intermediate ages and eventually become passive, old galaxies, like ellipticals, with similar metallicities.

\item The evolution of galaxies in A1314 can be perceived in three different sections on the basis of their location in the cluster, but all following the same processes of initially forming stars early in the Universe, assembling in the cluster, tidally stripped off their gas by ram pressure, turn passive and later merge with other stellar populations. The first section ($\leq$ 200 kpc), hosts the very old, massive populations (of red and dead galaxy types) with metallicities of solar values, evolving passively with mass accumulation by dry, minor mergers. The second set of red sequence galaxies, between 200-500 kpc, are tidally stripped galaxies (or in the process of being tidally stripped) by the interaction with the ICM; these galaxies display moderate star formation history with low to solar metallicities. The third set of population ($>$ 500 kpc) of mostly transition types with substantial star formation history and ages of 8.7-11.5 Gyr with a range of metallicity (0.2 to -1.2) and mass (-16.5 to -20.3 mag). This region at the edge of the cluster hosts the incoming field galaxies that maybe hybrid objects of pseudo bulges (or blue sequence E/S0) and blue Butcher-Oemler like galaxies.

\item During the course of this analysis, some hybrid red sequence systems were found (labelled in Table 2), which needs further inspection with deeper photometric surveys at different wavelengths. 
\end{enumerate}

\section*{Acknowledgments}

I would like to thank (late) Karl Rakos and Andrew Paul Odell for their immense support, guidance and assistance over the years, for which I am very grateful to them. I also thank Gerhard Hensler and Werner Zeilinger for their helpful suggestions, inspiration, and an anonymous referee for his useful comments to improve this paper.

%\bsp

\label{lastpage}

\end{document}